# A Modified GHG Intensity Indicator: Toward a Sustainable Global Economy based on a Carbon Border Tax and Emissions Trading

Reza Farrahi Moghaddam, Fereydoun Farrahi Moghaddam and Mohamed Cheriet

*Synchromedia Laboratory for Multimedia Communication in Telepresence,*
*École de technologie supérieure, Montreal (Quebec), Canada H3C 1K3*
*Tel.: +1(514)396-8972*
*Fax: +1(514)396-8595*
*rfarrahi@synchromedia.ca, imriss@ieee.org, ffarrahi@synchromedia.ca, mohamed.cheriet@etsmtl.ca*

**Abstract**

It will be difficult to gain the agreement of all the actors on any proposal for climate change management, if universality and fairness are not considered. In this work, a universal measure of emissions to be applied at the international level is proposed, based on a modification of the Greenhouse Gas Intensity (GHG-INT) measure. It is hoped that the generality and low administrative cost of this measure, which we call the Modified Greenhouse Gas Intensity measure (MGHG-INT), will eliminate any need to classify nations. The core of the MGHG-INT is what we call the IHDI-adjusted Gross Domestic Product (IDHIGDP), based on the Inequality-adjusted Human Development Index (IHDI). The IDHIGDP makes it possible to propose universal measures, such as MGHG-INT. We also propose a carbon border tax applicable at national borders, based on MGHG-INT and IDHIGDP. This carbon tax is supported by a proposed global Emissions Trading System (ETS). The proposed carbon tax is analyzed in a short-term scenario, where it is shown that it can result in significant reduction in global emissions while keeping the economy growing at a positive rate. In addition to annual GHG emissions, cumulative GHG emissions over two decades are considered with almost the same results.

## 1. Introduction

The Kyoto Protocol can be considered the foremost international agreement on climate change. However, some countries have withdrawn from the Protocol, or will withdraw from it, in spite of their initial support. The Kyoto Protocol is based on the hypothesis that a nation's economy is independent of the economies of other nations. This hypothesis is a reflection of the macro-canonical approach of the protocol's designers, and led to a commitment to reduce emissions to 5.2% percent below 1990 levels between 2008 and 2012. However, in our highly interconnected global economy and with new economic heavy-weights now on the scene, a more pragmatic approach is required to achieve a strategic goal such as the reduction of global GHG emissions [1, 2, 3, 4, 5, 6, 7, 8, 9, 10, 11, 12, 13, 14, 15, 16, 17, 18].

It is worth noting that, despite a general tendency to blame $CO_2$ emissions on non-green activities, any human footprint in nature, such as fossil fuel consumption (heat generation), Greenhouse Gas (GHG) emissions, anthropogenic activities (deforestation and urbanization), and water usage, could be considered as a non-green act. That said, $CO_2$ and GHG emissions still have a major impact on the environment and on the earth's atmospheric stability, and are the focus of this work.

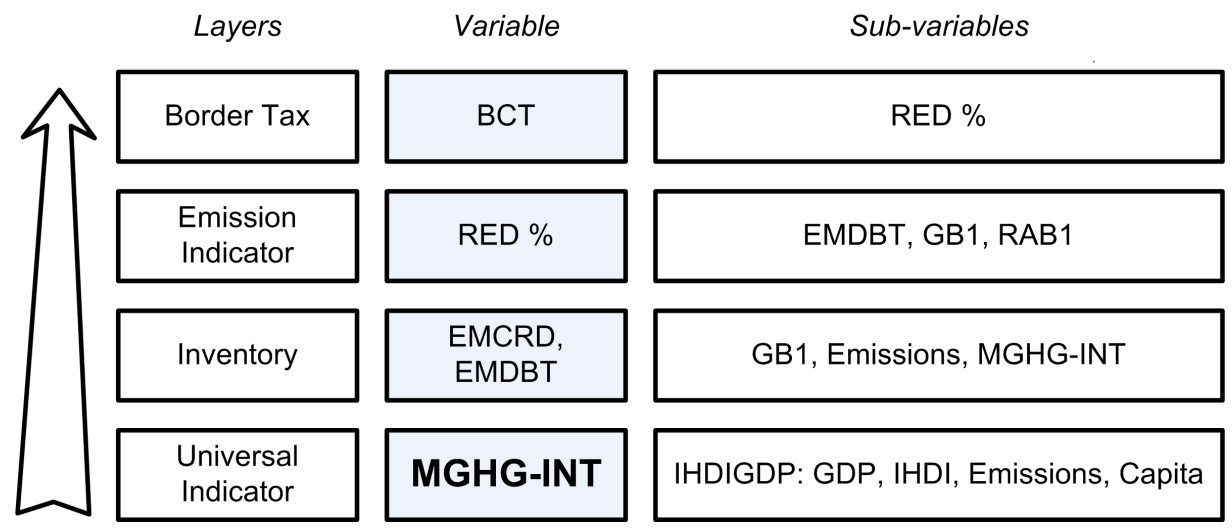

Figure 1: The proposed framework which is based on a universal indicator and consists of several layers. The details are presented in section 2 and subsequent sections.

---

Although several emissions indicators have been devised, such as greenhouse gas intensity (GHG-INT) and emissions per capita, there is no global agreement on any of them, as such measures can be very well received in one country and highly unpopular in others.

In this work, a universal indicator and measure of emissions is introduced to resolve the issues of world-level dissolution and lack of agreement. This indicator, which we call the Modified GHG Intensity (MGHG-INT) measure, not only takes into account the global influence of a nation in terms of its productivity (external "activity"), but also attempts to include its heretofore ignored internal "activity" to arrive at a universal emissions measure which can be used to evaluate a nation's contribution to the unsustainability of the planet, and also to react to it.

The absence of a universal indicator has resulted in the exclusion of developing countries from the Kyoto Protocol (and the same seems to be true of the successor agreements, like the Cancun Agreements and others). Not only has this resulted, and will result in high emissions leakage to developing countries, but it also causes other countries to withdraw these mechanisms. In contrast, mechanisms built using our proposed universal indicator, or any other similar universal indicator, have the advantage of treating all countries equally and fairly. With their minimal administrative costs, these mechanisms could put achievable goals within the grasp of all nations, and give the policymakers of each country the freedom to draw their own roadmap, and the challenge of doing so, to reduce the border tax or border tax adjustments their nation faces.

The schematic of the framework is presented in Figure 1. Its core is its universal indicator, the MGHG-INT measure. On the next layer, an international-level inventory is considered, which accounts for the credits and debits of nations based on the MGHG-INT indicator. On the third layer, an emission indicator is defined which accounts for the non-greenness of a nation's emissions. Finally, on the top layer, a carbon border tax is placed to empower nations to react unilaterally to other nations' non-green emissions even in the lack of an international agreement.

Despite the commonly held belief that World Trade Organization (WTO) regulations prevent the imposition of a carbon border tax, several groups are poring over World Trade Agreement (WTA) articles to find a way to implement such a tax [19, 20, 21, 22, 23, 24]. A border tax is a direct tax on imported goods, while a border tax adjustment (BTA) involves the imposition of a domestically imposed excise tax on "like" imported goods that are not sustainably produced [25]. The United States and the European countries in particular are working toward a border tax adjustment, which may also cover carbon leakage [23, 26, 27, 28]. In a recent special issue of Climate Policy, entitled, "Consuming and producing carbon: what is the role for border measures?", various aspects of border carbon adjustments and their role in preserving industrial competitiveness and preventing carbon leakage have been discussed in the context of world trade law [26, 27, 28]. There are many possible options in the WTO regulations that can be used to impose a border tax or BTA to control and reduce global GHG emissions. For example, Article II: 2(a) of the General Agreement on Tariffs and Trade (GATT) allows members of the WTO to place a border tax on the importation of any good, equivalent to an internal tax on a "like", or similar, good, and Article III: 2 of the GATT states that a BTA cannot be applied in excess of that applied directly or indirectly on a similar domestic good. If we read the wording of these articles in reverse, we find that any carbon border tax is possible, as long as an equivalent internal tax is implemented. The equivalent internal tax mechanisms are beyond the scope of this work, as we focus here solely on the carbon border tax. The most important aspect of border taxes is trade neutrality, i.e., the impact of border taxes on imported goods should be the same as the impact of internal tax mechanisms, or taxes imposed on domestic goods. [19, 20]. It is worth noting that our goal in this work is not to design a border policy based on border carbon adjustments, but to show that our universal indicator can be implemented in the form of a border policy mechanism. Also, some work has been conducted on applying uniform carbon taxes and analyzing their impact on the carbon footprint and economy in different sectors [29]. However, they did not consider the contribution of the country of origin to the global carbon footprint, but only the emissions related to the production of a good itself.

Usually, $CO_2$ emissions come from three sources: for energy consumption and food production, emitting in the concrete industry, and emitting as the result of land use. The proposed carbon tax manages the emissions related to energy consumption and food production, while the proposed ETS, which provides a means for emissions trading to reduce carbon taxes, covers land-use emissions. The details of this tax are provided in the following sections. In brief, a carbon tax mechanism to be implemented at a country's border as a border tax or a border tax adjustment is proposed, which consists of two terms:

Carbon tax of origin + Inter-country Transportation carbon tax

Because of complexity of transportation mechanisms across international borders, this term has been separated from the carbon tax of origin. In this work, we only consider the carbon tax of origin. In future, and using a multi-region input-output (MRIO) model, another framework will be proposed to estimate the transportation carbon tax. The estimated total emissions of the international shipping transport sector, which is responsible for 80% of the global trade in goods, is estimated to represent 2.7% of the world's $CO_2$ emissions from fossil fuel combustion [30]. However, because of the highly heterogeneous nature of international shipping activities and also the high level of heavy fuel oil consumption by this sector, control mechanisms are very important in order to prevent the production of high levels of hidden emissions by these activities.

The paper is organized as follows. In section 2, the proposed framework is described at high level. The definitions of IDHIGDP and MGHG-INT are presented in section 3. The RED percentage is defined in section 4, and used to define the proposed carbon tax in section 5. A simulation of the impact of a carbon border tax on the world economy is presented in section 5.1. The proposed ETS is provided in section 5.2. Finally, our concluding remarks and future research prospects are presented and discussed in section 6. In section A of the appendix, common notations and definitions used in other sections are introduced. The motivations for the new emissions indicator is presented in greater detail in section B. The issue of the accumulation of GHG emissions over two decades is discussed in section C of the appendix. Also, full-size tables of various variables and indicators are provided as supplementary material.

## 2. The proposed framework

Figure 2 shows the status of the world in terms of GDP and emissions in 2009. While there have been some fluctuations, the overall picture remains the same today, with the presence of two heavy-weights, the United States and China. One would think, therefore, that control and penalty measures designed to reduce the human footprint on nature will have the same distribution across the world.

However, traditional measures, such as GHG Emission Intensity (GHG-INT) and GHG emissions per capita, penalize only one of the two major emitters, ignoring the other one (see Figure 2). The reason for this could be related to differences between these two countries in terms of population, wealth distribution, quality of life, culture, legal philosophy, political power distribution, corporate governance, and regulatory framework, for example. However, all countries evolve over time, and proposing a measure based on the current status of a country is not recommended, as it may prove ineffective (or even have an unintended effect) in the future. Our objective is to provide a measure which covers all emitters uniformly, despite their differences. This new measure, shown pictorially in Figure 3, should be stable over time, high-level, and easy to calculate. We call it the Modified GHG Intensity (MGHG-INT), as we think it should calculate performance with respect to "activity". The details of this proposed measure are provided in section 3 based on an IHDI-adjusted gross domestic product, or IDHIGDP.

There are several global threats that support the urgent need for a new universal measure: i) global disagreement on emission reduction goals, measures, and procedures, ii) de-industrialization of Europe, iii) leakage and hidden emissions in exports/imports and transport, iv) population, and v) administration cost. We discuss them in details in Appendix B.

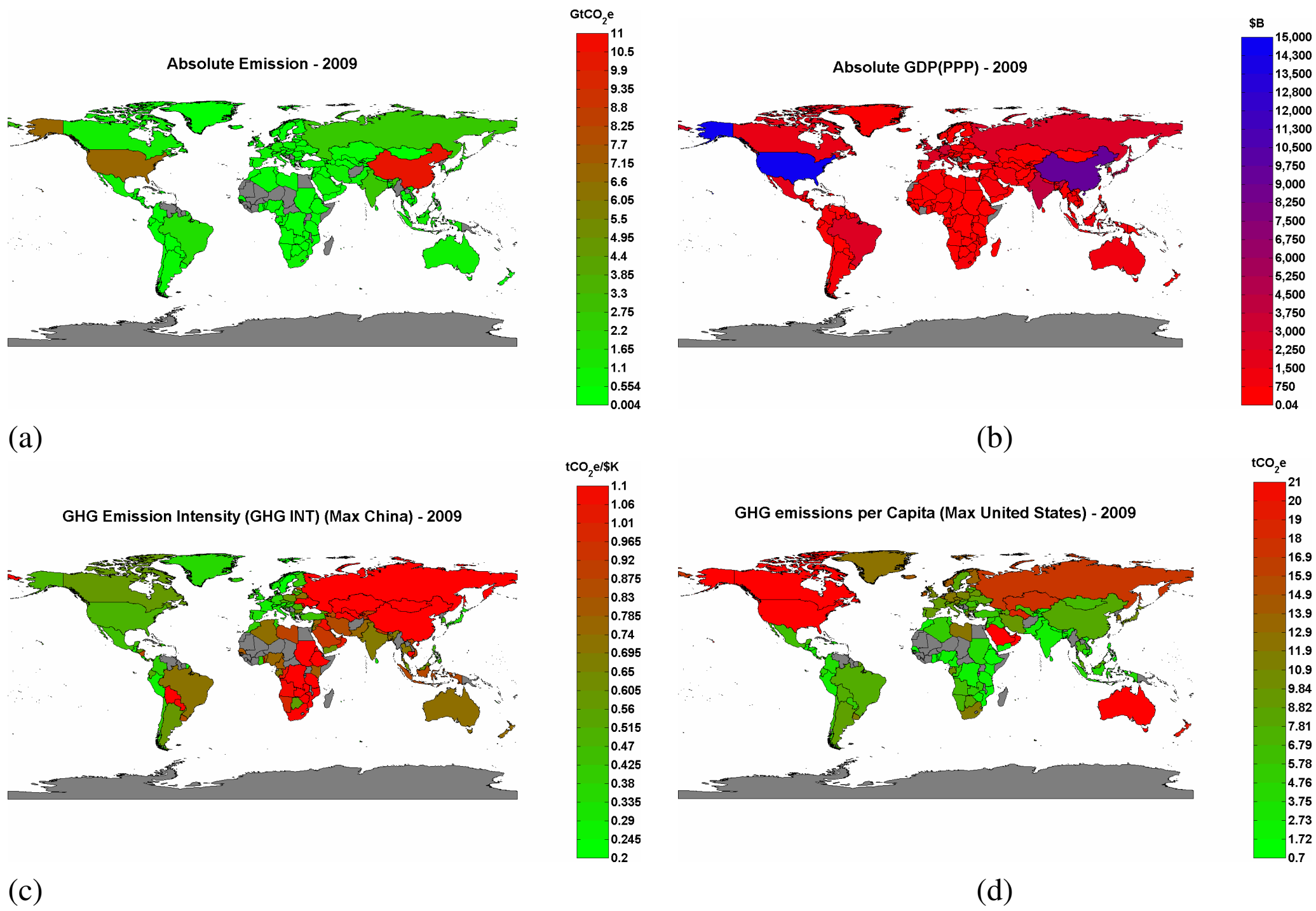

Figure 2: a) Global picture of GHG emissions (in gigatonnes of carbon dioxide equivalent ($GtCO_2e$)) in 2009. b) Global distribution of GDP (PPP) in the same year (in billions of international dollars). c) GHG Emission intensity (GHG INT) in 2009 (in $GtCO_2e$/\$B). d) GHG emissions per capita in the same year (in $GtCO_2e$/Million Capita). *Data Sources:* US Energy Information Administration, World Bank, United Nations Statistics and Research Database, International Monetary Fund, and United Nations Development Programme. For more details, please see section 2.2.

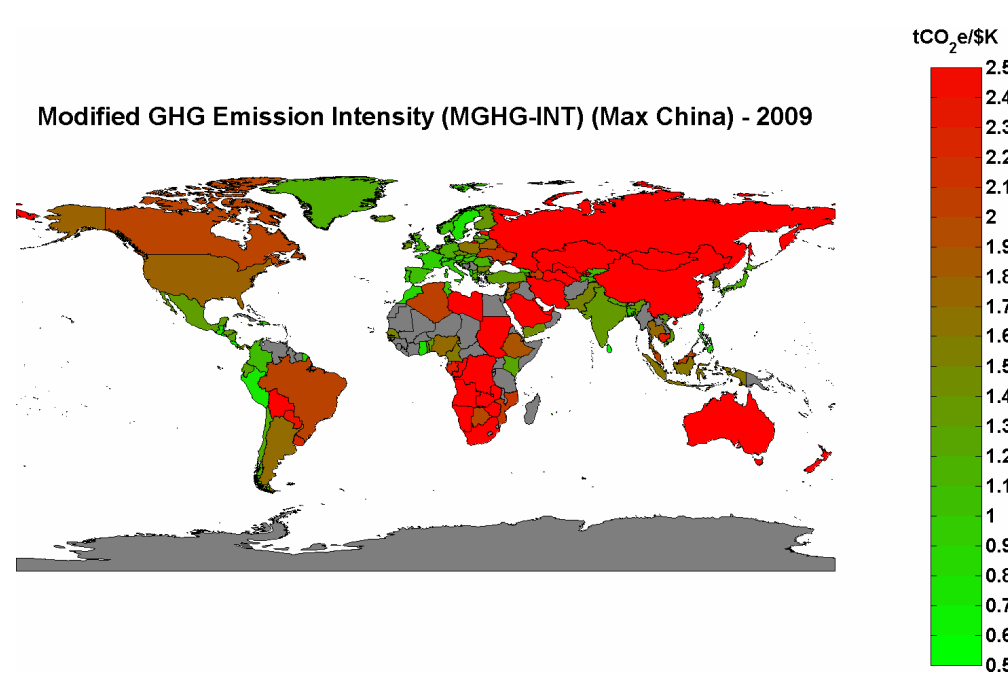

Figure 3: What an ideal indicator of emissions should look like, and which the MGHG-INT, introduced in section 3, will provide. The details are provided in section 3.

It is worth noting that the main contribution of this paper is the introduction of the MGHG-INT indicator as a universal indicator of emission intensity for all countries, and so no other form of classification, into developed/developing countries, for example, is needed. From the standpoint of a universal indicator, a tonne of $CO_2$ does not always mean the same as a tonne of $CO_2$ in terms of "activity". Therefore, we propose a new inventory to account for the emission credits and debits of nations based on their activities, which can be used in control mechanisms (such as a border carbon tax) designed to help move the world toward a sustainable future. In order to show the applicability of this indicator, the national level emission credits and debits are first defined, and a trading system, powered by the RED percentage indicator, is introduced. Then, an enforcing mechanism at the international level is introduced by defining the international level border carbon taxes (BCTs) or adjustments (BCAs) based on the RED percentages. This mechanism is assumed to be backed up by an internal equivalent carbon policy in each country, who wishes to impose CBAs on imports from other countries. The main advantages of the proposed framework are its simplicity and straightforwardness, which can help it gain trust and acceptance of everyone involved, and its low administration costs, because the emissions associated with a specific imported good are not required in the calculations. Furthermore, the unilateral nature of the framework enables any country to implement it without the approval of the others. It is worth noting that our enforcing mechanism developed based on the MGHG-INT, is not the only possible one, and the authors hope that the MGHG-INT, as a universal indicator of carbon intensity, could open up a new research area in universal approaches to the world's sustainable development.

### *2.1. Gross Domestic Product (GDP) and Purchasing Power Parity GDP (GDP (PPP))*

Gross Domestic Product (GDP), which was first developed by Simon Kuznets for a US Congress report in 1934 [31], is a measure of the market value of all final goods and services produced in a country in a given period. In our calculation, we use the GDP at purchasing power parity exchange rates (GDP (PPP)) as a measure of the activity level of a country, because it reflects living standards more accurately than exchange rates. The GDP (PPP) attempts to relate changes in the nominal exchange rate between two countries currencies to changes in those countries price levels [32]. We use the values calculated by the International Monetary Fund[1] in international dollars. The 5 countries with highest GDP (PPP) in 2009 are listed in Tables 1.

It is also worth noting that some alternatives to the GDP, such as the Index of Sustainable Economic Welfare (ISEW) and the Genuine Progress Indicator (GPI), which have been proposed to augment it with considerations such as social and environmental aspects, and economic costs and benefits [33, 34, 35, 36, 37, 38]. In this work, we use the GDP (PPP) index, which is well accepted, and hybridize it with the IHDI, which results in the IHDI-adjusted GDP (IHDIGDP) (introduced in section 3).

| **Rank** | **Country** | **GDP (PPP)** |
|---|---|---|
| **1** | **United States** | **14,120** |
| **2** | **China** | **9,057** |
| **3** | **Japan** | **4,107** |
| **4** | **India** | **3,645** |
| **5** | **Germany** | **2,814** |

Table 1: Top 5 countries in terms of GDP (PPP) (in $B) in 2009.

| **Rank** | **Country** | **Emissions (MtCO2e)** |
|---|---|---|
| **1** | **China** | **10,060** |
| **2** | **United States** | **6,581** |
| **3** | **India** | **2,432** |
| **4** | **Russia** | **2,361** |
| **5** | **Brazil** | **1,385** |

Table 2: Top 5 countries in terms of GHG emissions in 2009.

[1] http://www.imf.org/external/pubs/ft/weo/2011/01/weodata/index.aspx

### *2.2. Data*

In this work, we use the $CO_2$ and non-$CO_2$ emissions data over a period of 30 years from 1980 to 2009. These data were obtained from the US Energy Information Administration and World Bank databases [39]. The economic and social indicators, such as GDP at purchasing power parity exchange rates (PPP) [40], population, and the IHDI [41, 42], were obtained from the United Nations Statistics and Research Database (UNdata), the International Monetary Fund (IMF), and the United Nations Development Programme (UNDP) database covering the same period of time. The Green and Red scenarios, which will be referred to in section 4, were built based on the B1 Asian-Pacific Integrated Model (AIM) and the A1B AIM scenarios of the Intergovernmental Panel on Climate Change (IPCC) [43, 44, 45, 46, 47, 48, 49, 50, 51, 52, 53, 54] referred to in section A.3 of the appendix. More information on the data and the source links are given in the notation section (section A of the appendix).

The use of names of countries in the figures and tables serves only to identify world regions, and does not imply the expression of any opinion on the legal status of any country or its authorities, or concerning its boundaries.

## 3. IHDI-adjusted Gross Domestic Product (IHDIGDP) and Modified GHG Emission Intensity (MGHG-INT)

The GHG-INT of a country is defined as the ratio of its emissions to its GDP:

$$\mathrm{GHGINT}_{i,y} = \frac{\mathrm{EM}_{i,y}}{\mathrm{GDP}_{i,y}} \tag{1}$$

where $\mathrm{GHGINT}_{i,y}$ is the GHG-INT of country $i$ in year $y$, and $\mathrm{EM}_{i,y}$ is the total emissions of that country in the same year (excluding land-use emissions). This measure provides the GHG footprint of countries based on their economic output. If the GHG emissions of a country are proportional to its GDP, then this measure is low for that country. In contrast, if the amount of GHG emissions is relatively higher than the GDP of a country, then that country is designated a red emissions zone according to this measure. For example, if a less productive country with a small GDP is producing a large amount of GHG gases, then its GHG-INT is not acceptable. According to the GHG-INT measure, countries like the United States with very high GHG emissions and a high GDP are safely in the green zone, while countries like China with high GHG emissions, but not a correspondingly high GDP are in the red zone. Of course, countries like China with a large population prefer to use a different measure, which is a ratio of emissions to the population (GHGpCapita):

$$\mathrm{GHGpCapita}_{i,y} = \frac{\mathrm{EM}_{i,y}}{\mathrm{Capita}_{i,y}} \tag{2}$$

where $\mathrm{GHGpCapita}_{i,y}$ is the GHGpCapita of country $i$ in year $y$, and $\mathrm{Capita}_{i,y}$ is its population. This measure represents the GHG footprint of a country based on its population. For a country with GHG emissions that are proportional to its population, GHGpCapita is an acceptable measure, but if the amount of its GHG emissions is relatively more than the size of its population, then that country is in the red zone. For example, if a small country with a small population produces a large amount of GHG gases, it will not find the GHGpCapita an attractive measure. According to the GHGpCapita, countries like China, with very high GHG emissions and a large population, are in the green zone. In contrast, countries like the United States, with high GHG emissions but a relatively small population, are in the red zone.

To arrive at a universal GHG emissions measure which is robust with respect to variations in GDP and population, but works for all countries, we modify the GHG-INT, and redefine it as the ratio of emissions to "activities":

$$\mathrm{MGHGINT}_{i,y} = \frac{\mathrm{EM}_{i,y}}{\text{“activities”}_{i,y}} \tag{3}$$

where $\mathrm{MGHGINT}_{i,y}$ is the modified GHG intensity measure of country $i$ in year $y$ (defined above), and $\text{“activities”}_{i,y}$ is the activity of that country (explained below) during the same period. Here, "activities" replaces GDP in Equation (1). We model them as an IHDI-adjusted version of GDP (IHDIGDP), which not only includes the production of a country (its GDP), but also considers the internal activity of its population. We formally introduce the IHDIGDP later in this work.

Using the IHDIGDP, we redefine MGHG-INT as follows:

$$\mathrm{MGHGINT}_{i,y} = \frac{\mathrm{EM}_{i,y}}{\mathrm{IHDIGDP}_{i,y}} \tag{4}$$

where $\mathrm{EM}_{i,y}$ represents the total GHG emissions of that country, except for the land-use $CO_2$ emissions.

| Rank | Country | IHDI |
|---|---|---|
| 1 | Norway | 8,751 |
| **2** | **Australia** | **8,622** |
| 3 | Sweden | 8,231 |
| 4 | The Netherlands | 8,162 |
| **5** | **Germany** | **8,122** |

(a)

| COI | IHDI | COI | IHDI |
|---|---|---|---|
| **Australia** | **8,622** | **Indonesia** | **4,882** |
| **Brazil** | **5,046** | **Japan** | **4,808** |
| **Canada** | **8,103** | **Russia** | **6,316** |
| **China** | **5,048** | **South Africa** | **4,089** |
| **Germany** | **8,122** | **Switzerland** | **8,111** |
| **India** | **3,601** | **United States** | **7,963** |

(b)

Table 3: The IHDI (in a scale of 1,000) in 2009: a) for the top 5 countries, b) for the countries of interest (COI). The COI is composed of countries with global or regional influence, and includes *Australia, Brazil, China, Canada, Germany, India, Indonesia, Japan, Russia, South Africa, the United States, and Switzerland*.

---

| Rank | Country | IHDIxCapita |
|---|---|---|
| **1** | **China** | **5,772,000** |
| **2** | **India** | **3,104,000** |
| **3** | **United States** | **1,991,000** |
| **4** | **Russia** | **932,800** |
| **5** | **Indonesia** | **878,000** |

(a)

| COI | IHDIxCapita | COI | IHDIxCapita |
|---|---|---|---|
| **Australia** | **148,000** | **Indonesia** | **878,000** |
| **Brazil** | **739,800** | **Japan** | **593,500** |
| **Canada** | **223,900** | **Russia** | **932,800** |
| **China** | **5,772,000** | **South Africa** | **150,700** |
| **Germany** | **640,700** | **Switzerland** | **54,440** |
| **India** | **3,104,000** | **United States** | **1,991,000** |

(b)

Table 4: The IHDIxCapita ($\times 10^6$) in 2009: a) for the top 5 countries, b) for COI. Note: The 1900 population snapshots were used in the calculation of IHDIxCapita

---

Let us start first with IHDIxCapita. The IHDIxCapita is defined as the product of the UN Development Programme's IHDI and the population snapshot (see section A.2 for further detail):

$$\mathrm{IHDIxCapita}_{i,y} = \mathrm{IHDI}_{i,y}\mathrm{Capita}_{i,1990} \tag{5}$$

where $\mathrm{IHDI}_{i,y}$ is the IHDI of country $i$ in year $y$, $\mathrm{Capita}_{i,1990}$ is population snapshot of country $i$ (taken in 1990), and $\mathrm{IHDIxCapita}_{i,y}$ is the IHDIxCapita of the same country in year $y$. The countries with highest IHDI and IHDIxCapita are listed in Tables 3 and 4. The balanced IHDIxCapita is defined as the IHDIxCapita normalized to the maximum IHDIxCapita in the same year, scaled to the maximum GDP (PPP) of the same year:

$$\mathrm{IHDIxCapita}^{\mathrm{BAL}}_{i,y} = \mathrm{GDP(PPP)}^{\mathrm{MAX}}_{y} \frac{\mathrm{IHDIxCapita}_{i,y}}{\mathrm{IHDIxCapita}^{\mathrm{MAX}}_{y}} \tag{6}$$

where $\mathrm{IHDIxCapita}_{i,y}$ is the IHDIxCapita of country $i$ in year $y$, and $\mathrm{IHDIxCapita}^{\mathrm{MAX}}_{y}$ and $\mathrm{GDP\ (PPP)}^{\mathrm{MAX}}_{y}$ are the maximum of the IHDIxCapita and the maximum of GDP (PPP) of all countries in year $y$ respectively. The balanced GDP, $\mathrm{GDP}^{\mathrm{BAL}}_{i,y}$, is the ratio of the GDP (PPP) to the GDP (PPP) of the country with the IHDI of $\mathrm{IHDIxCapita}^{\mathrm{MAX}}_{y}$, scaled to the maximum GDP of the same year:

$$\mathrm{GDP}^{\mathrm{BAL}}_{i,y} = \mathrm{GDP(PPP)}^{\mathrm{MAX}}_{y} \frac{\mathrm{GDP\ (PPP)}_{i,y}}{\mathrm{GDP\ (PPP)}^{\mathrm{IHDI}}_{y}} \tag{7}$$

where $\mathrm{GDP\ (PPP)}^{\mathrm{IHDI}}_{y}$ is the GDP (PPP) of the country with the IHDI of $\mathrm{IHDIxCapita}^{\mathrm{MAX}}_{y}$.

With this definition, if we calculate the balanced GDP of the country with the maximum IHDIxCapita, we obtain the GDP (PPP) of the country with the maximum GDP. Also, the balanced IHDIxCapita of a country with the maximum IHDIxCapita is again the GDP (PPP) of the country with the maximum GDP. In this way, both the balanced

GDP and the balanced IHDIxCapita are normalized to the same level, and therefore it is possible to average them and calculate the IHDIGDP. The IHDIGDP is defined as follows:

$$\mathrm{IHDIGDP}_{i,y} = Z\frac{\mathrm{GDP}_{i,y}^{\mathrm{BAL}} + \mathrm{IHDIxCapita}_{i,y}^{\mathrm{BAL}}}{2} \tag{8}$$

where $\mathrm{IHDIGDP}_{i,y}$ is the IHDIGDP of country $i$ in year $y$, and $\mathrm{IHDIxCapita}_{i,y}^{\mathrm{BAL}}$ and $\mathrm{GDP}_{i,y}^{\mathrm{BAL}}$ are the balanced IHDIx-Capita and the balanced GDP of that country in year $y$ respectively. The normalization parameter $Z$ is selected is such a way that the world IHDIGDP in 1990 is equal to the world GDP (PPP) in the same year.

If a country has a GDP (PPP) higher than GDP $(\mathrm{PPP})_y^{\mathrm{IHDI}}$, it will have a $\mathrm{GDP}^{\mathrm{BAL}}$ higher than the GDP $(\mathrm{PPP})_y^{\mathrm{MAX}}$. At the same time, the balanced IHDIxCapita is the ratio of the IHDIxCapita to the maximum IHDIxCapita, scaled to GDP $(\mathrm{PPP})_y^{\mathrm{MAX}}$. Every country has an IHDIxCapita lower than the maximum IHDIxCapita, and therefore it will have a balanced IHDIxCapita that is lower than GDP $(\mathrm{PPP})_y^{\mathrm{MAX}}$.

The advantage of the IHDIGDP over the GDP is that countries with a large population and those with a high GDP are treated the same way, and benefit from this measure. For countries with high GDPs, the balanced GDP is high, and for countries with a large population and a good IHDI factor, the balanced IHDIxCapita is high. In both cases, the IHDIGDP is high, and therefore their footprint, MGHG-INT, will be small. But, if a country suffers from with a low GDP, or a large population but low quality of life, then both $\mathrm{GDP}^{\mathrm{BAL}}$ and $\mathrm{IHDIxCapita}^{\mathrm{BAL}}$ will be low, which results in a high MGHG-INT value, and indicates that the emission performance of that country should be improved.

The countries with the highest IHDIGDP values are listed in Table 5. These values are compared with those of countries with the highest GDP. We can conclude that the IHDIGDP is a better measure, as it gives a more balanced picture of a country's activities status. The distributions of the IHDIGDP and the GDP per capita are illustrated in Figures 4(a) and 4(b) respectively.

| Rank | Country | IHDIGDP |
|---|---|---|
| 1 | China | 4,026 |
| 2 | United States | 3,832 |
| 3 | India | 1,892 |
| 4 | Japan | 1,120 |
| 5 | Germany | 848.8 |

(a)

| Rank | Country | GDP (PPP) |
|---|---|---|
| 1 | United States | 14,120 |
| 2 | China | 9,057 |
| 3 | Japan | 4,107 |
| 4 | India | 3,645 |
| 5 | Germany | 2,814 |

(b)

Table 5: a) The 5 countries with the highest IHDIGDP in 2009 (in billions of dollars). b) The 5 countries with the highest GDP in the same year (in billions of international dollars).

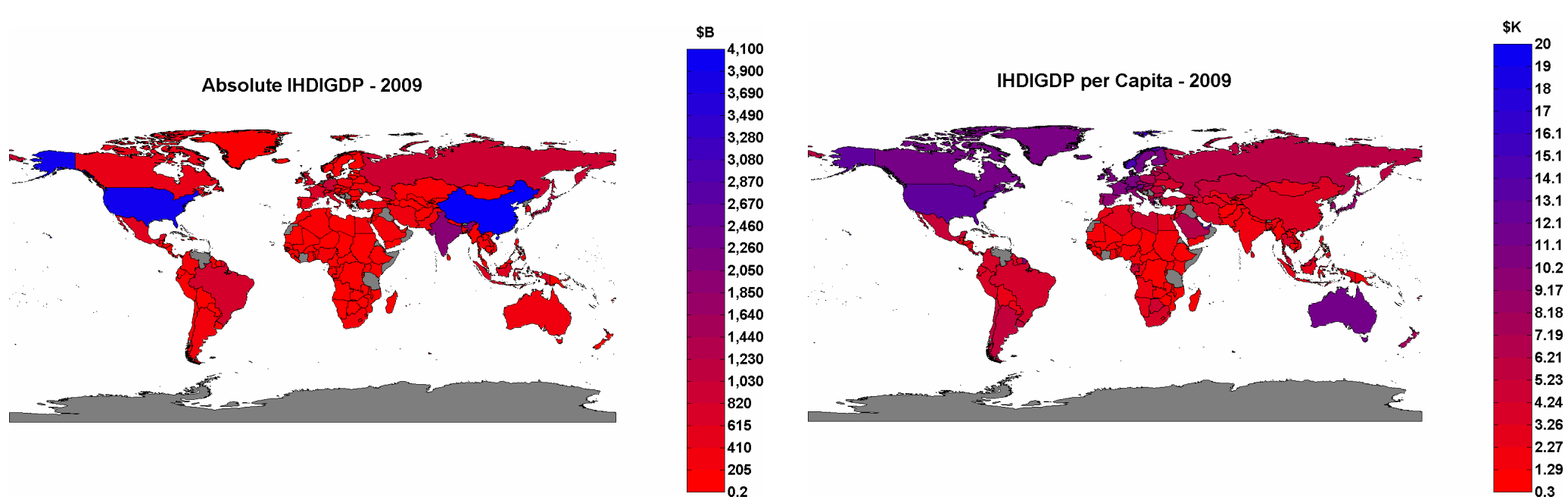

(a) (b)

Figure 4: a) The IHDIGDP distribution over the world in 2009 (in billions of dollars). b) The IHDIGDP per capita in the same year (in thousands of dollars).

It is interesting to compare the variations in GDP (PPP) and IHDIGDP variation over time. Figure 5 shows the profiles of GDP (PPP) and IHDIGDP over two decades. It is worth noting that the IHDIGDP shows a very stable and smooth increase compared to the bubbling effect of the GDP (PPP). This stability can be put down to the fact that the IHDIGDP is normalized to the population activities that are less dependent on the market fluctuations. The global IHDIGDP has grown by 1.67% annually over the last 30 years. This suggests that it can also be used in other economic analyses as a robust measure.

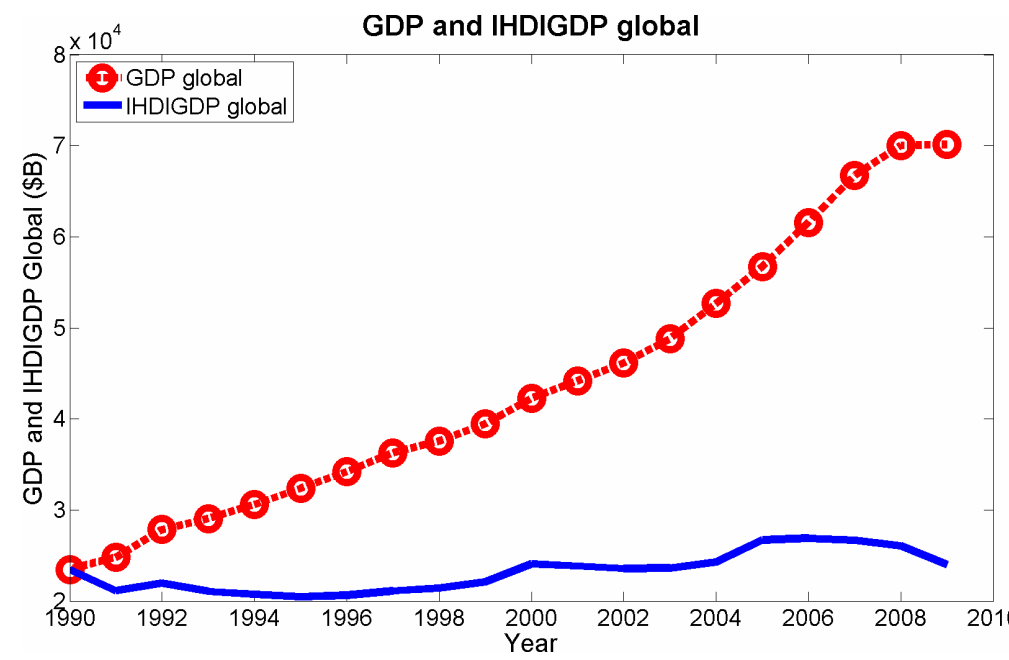

Figure 5: A comparison between global GDP and global IHDIGDP over two decades. The IHDIGDP shows very robust and stable growth.

---

Finally, a comparison of the various indicators for two major global players, the United States and China, is provided in Figure 6. In Figure 6(a), the variations in GHG-INT and MGHG-INT are compared. As we can see, GHG-INT and GHGpCapita are too far apart for either of the two countries to catch up with the other, and this has been resulted in disagreements in the past. However, the MGHG-INT values are close, and so the two countries could converge the values with a reasonable effort. In Figure 6(b), the GDP and IHDIGDP of the two countries are compared. Again, the IHDIGDP shows the competitiveness between the countries, and gives a better picture of their economies.

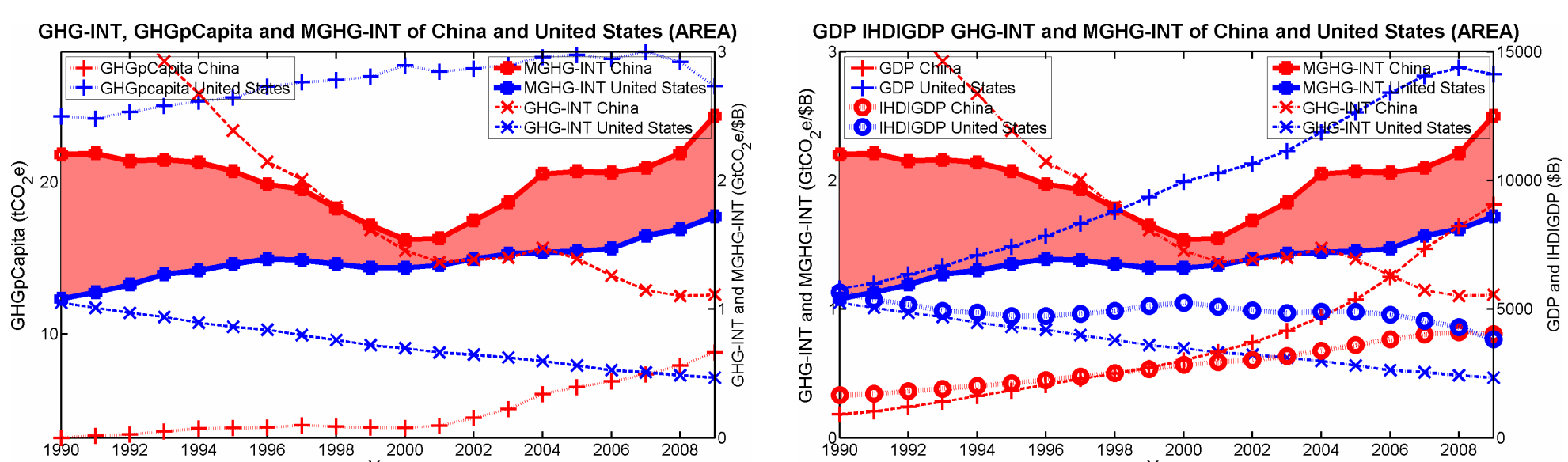

(a) (b)

Figure 6: a) A comparison between the GHG-INT, GHGpCapita, and MGHG-INT of China and the United States over two decades. b) The same comparison as in (a), but with respect to GDP, IHDIGDP, GHG-INT, and MGHG-INT.

---

It is worth noting that the IHDIGDP dollar is similar to the international GDP (PPP) dollar, but not the same. The IHDIGDP dollar cannot be exchanged with any currency, and serves only as a global and uniform measure for comparing all countries, regardless of their status in the world.

The global distribution of MGHG-INT is provided in Figure 7(a). As expected, all major emitters have a high MGHG-INT. The same distribution is shown in Figure 7(b), but limited to China's MGHG-INT, in order to provide a more detailed presentation. Figure 7(c) shows a comparison of the global MGHG-INT with the global GHG-INT over two decades. It is worth noting that the latter trends continuously downward, while the former remains almost constant, with a little increase. It can be argued that GHG-INT is not a good direct index of emissions reduction. In addition, it has been observed that the global MGHG-INT ($1,744$ $tCO_2e$/\$K in 2009) is lower than the MGHG-INT averaged across all nations ($2,400$ $tCO_2e$/\$K in 2009). This suggests that actually only a few administrations are carrying the global movement toward GHG reduction. The countries with highest MGHG-INT are also listed in Table 6.

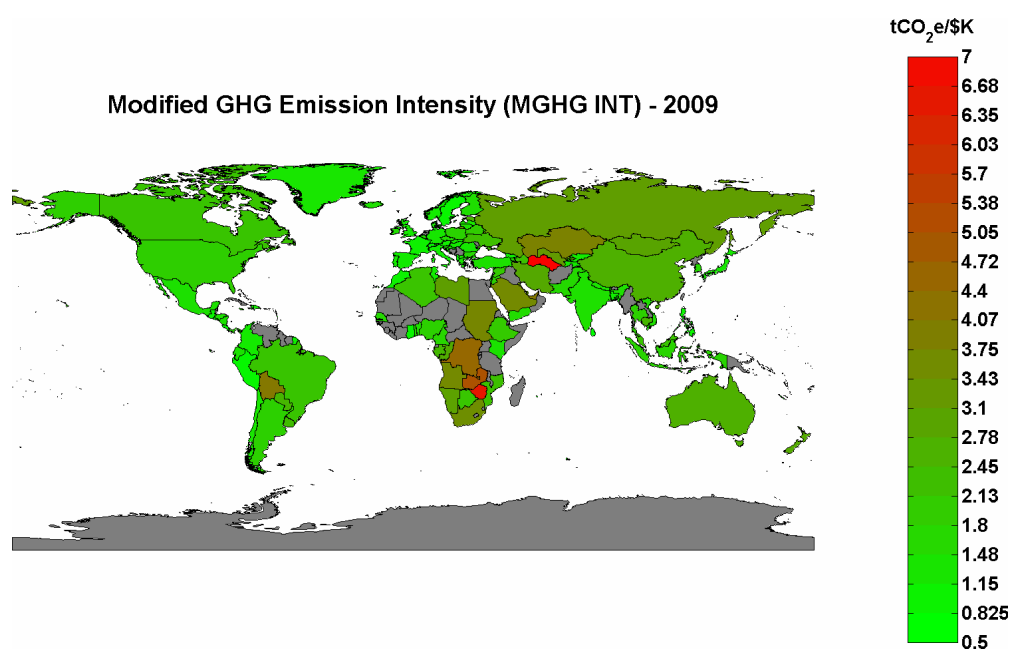

(a)

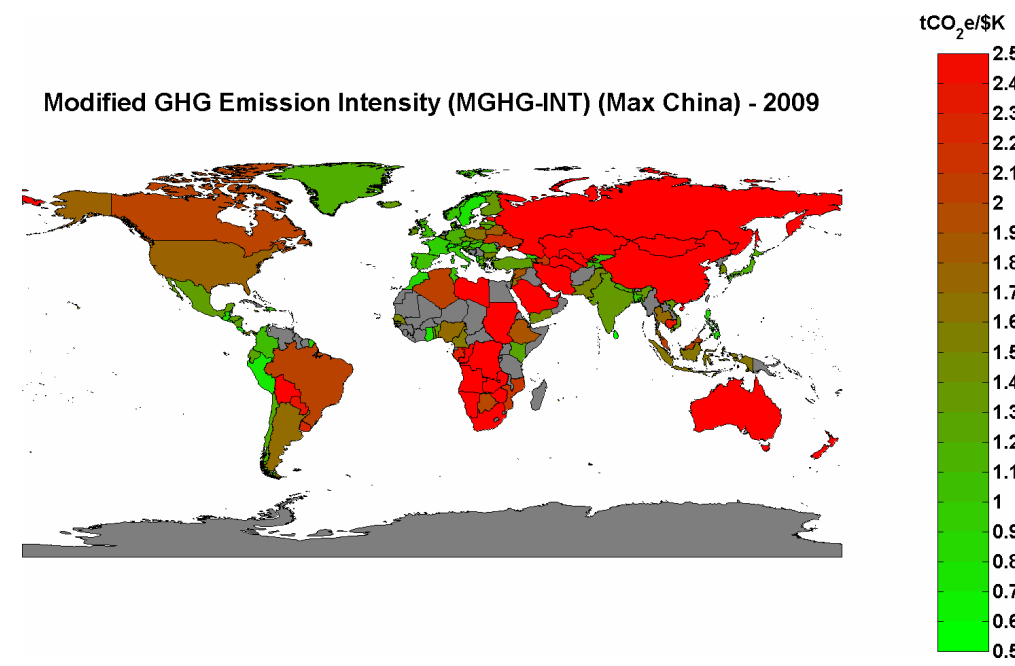

(b)

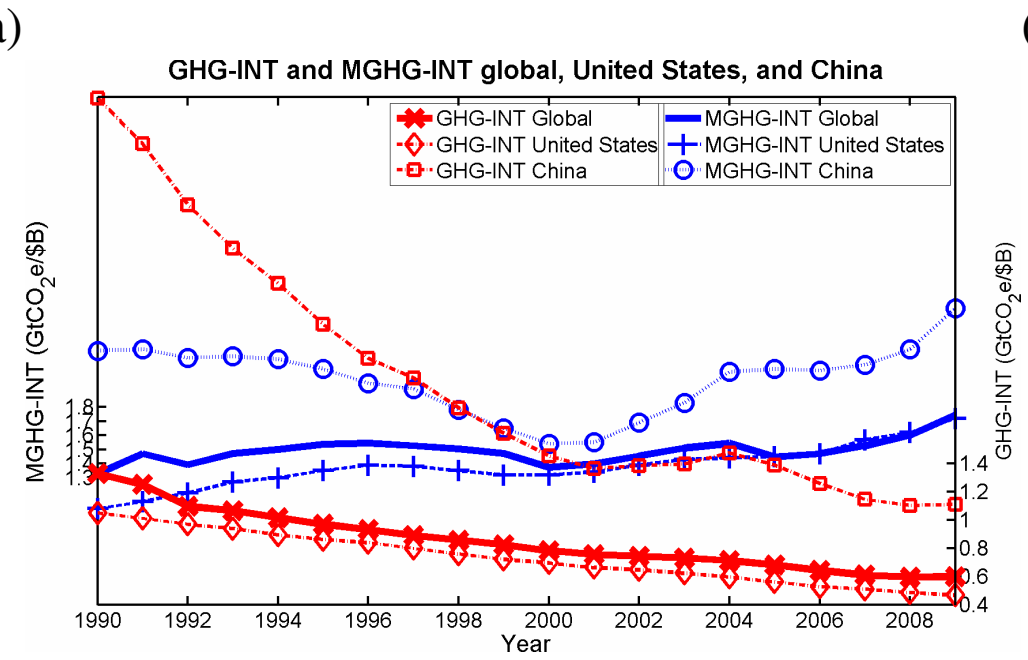

(c)

Figure 7: a) The MGHG-INT distribution over the world in 2009 (in $GtCO_2e$/\$B). b) The same as (a), but with the maximum value of China's MGHG-INT, in order to provide a clearer picture. c) The global MGHG-INT trend compared to the global GHG-INT trend over two decades. As discussed in the section on motivations, the MGHG-INT provides a universal measure of GHG emissions for all countries.

| **Rank** | **Country** | **MGHG-INT** |
|---|---|---|
| 1 | Turkmenistan | 6.91 |
| 2 | Zimbabwe | 6.4 |
| 3 | Bahrain | 4.92 |
| 4 | Zambia | 4.9 |
| 5 | Congo, DRC | 4.34 |

(a)

| COI | **MGHG-INT** | COI | **MGHG-INT** |
|---|---|---|---|
| **Australia** | **2.5** | **Indonesia** | **1.56** |
| **Brazil** | **1.97** | **Japan** | **1.11** |
| **Canada** | **1.96** | **Russia** | **2.97** |
| **China** | **2.5** | **South Africa** | **3.34** |
| **Germany** | **1.11** | **Switzerland** | **0.63** |
| **India** | **1.29** | **United States** | **1.72** |

(b)

Table 6: The MGHG-INT in 2009 (in $tCO_2e$/\$K): a) for the top 5 countries, b) for COI.

Although the MGHG-INT will provide a universal measure of the footprint potential at the international level, and can be used by investors to choose a region with higher efficiency and less risk of penalties, it cannot be used to directly calculate those penalties or the adjustments required to reduce emissions. The MGHG-INT concept is analogous to the power concept in a machine. The penalty should be applied based on the amount of energy the machine consumes, rather than on its power specifications. In the next section, the percentage of 'non-greenness' of the production of each nation is calculated using the IHDIGDP and MGHG-INT. This percentage will serve as the foundation for a carbon tax, which will be introduced in section 5.

## 4. RED Percentage: A Measure of Emission Inefficiency

In this section, we describe two GHG emission scenarios, Green and Red, representing safe (admissible emissions) and dangerous (over-emissions) levels for the world respectively. The Green emission scenario is based on the IPCC's B1 Asian-Pacific Integrated Model (AIM) scenario [43, 53] (see section A.3 for more details). We denote the Green scenario emissions in year $y$ by $\mathrm{GB1}_y$. We use the A1B AIM scenario as the base for building the Red scenario, denoted by $\mathrm{RA1B}_y$ (see section A.3). We distribute $\mathrm{GB1}_y$ among all countries based on their IHDIGDP, in order to calculate the green or B1 AIM share of each country:

$$\mathrm{ADMEM}_{i,y} = \frac{\mathrm{IHDIGDP}_{i,y}}{\mathrm{IHDIGDP}_y}\mathrm{GB1}_y \tag{9}$$

where $\mathrm{IHDIGDP}_{i,y}$ is the IHDIGDP of country $i$ in year $y$, $\mathrm{IHDIGDP}_y$ is the total IHDIGDP of the world, $\mathrm{GB1}_y$ is the global green admissible emissions in year $y$, and $\mathrm{ADMEM}_{i,y}$ is the admissible (B1 AIM share) emissions of country $i$ in year $y$. If the difference between the total emissions of a country and its green (B1 AIM) share is negative, then all its emissions are admissible, and the difference will be considered as the emission credits for that country:

$$\mathrm{EMCRD}_{i,y} = \begin{cases} -\left(\mathrm{EM}_{i,y} - \mathrm{ADMEM}_{i,y}\right) & \text{if } \mathrm{EM}_{i,y} - \mathrm{ADMEM}_{i,y} < 0 \\ 0 & \text{otherwise} \end{cases} \tag{10}$$

where $\mathrm{EMCRD}_{i,y}$ represents the emission credits of country $i$ in year $y$, and $\mathrm{EM}_{i,y}$ is the total emissions of that country in the same year. But, if the total emissions of a country are greater than its green share (its admissible emissions), the difference is the emission debt for that country:

$$\mathrm{EMDBT}_{i,y} = \begin{cases} \left(\mathrm{EM}_{i,y} - \mathrm{ADMEM}_{i,y}\right) & \text{if } \mathrm{EM}_{i,y} - \mathrm{ADMEM}_{i,y} > 0 \\ 0 & \text{otherwise} \end{cases} \tag{11}$$

where $\mathrm{EMDBT}_{i,y}$ is the emission debt of country $i$ in year $y$.

These concepts are illustrated in Figure 8. In Figure 8(a), the country's total emissions are less than its admissible emissions (ADMEM), and therefore it has some emission credits (EMCRD), which can be traded in the proposed ETS (see section 5.2). In contrast, in Figure 8(b), the country is in emission debt, because its total emissions are more than that of its B1 AIM share. In this case, a carbon tax will be imposed on the goods originating from that country (see section 5.1), except in the case that country clears its emission debt in the ETS. In Figures 9(a) and (b), the admissible emissions and emission debts are illustrated along with the Green and Red scenarios over short-term and long-term periods of time.

In Table 7, the 5 countries with the highest admissible emissions and lowest admissible emissions per capita are listed. The countries with lowest admissible emissions per capita are mostly located in Africa. In Table 8(a) and Table 8(b), the 5 countries with the highest emission debt and the highest emission credits in 2009 are listed respectively. Compared to 1990 ranking (Table 8(c)), we can see that emissions credits have been reduced mostly because of the emergence of new actors in the global economy. Interestingly, the USA, with a large proportion of the emissions credits in 1990, not only had no more credits in 2009, but also suffered an emissions debt in that year (cf. Table 8(a)). In a simplified imaginary analysis, it can be seen that the emissions credits of a European country, for example the UK, would be higher if it were not reducing its emissions. In fact, the UK emissions credits would be 356 $\mathrm{MtCO_2e}$, which is 60.8 $\mathrm{MtCO_2e}$ more than its actual emissions credits in 2009. In this analysis, we assumed that the only modified parameter is the UK GHG emissions in 2009, which was set to that of the UK in 1990: 780 $\mathrm{MtCO_2e}$. This leads to an increase in the IHDIGDP of the UK from $624.9 billion to $772.3 billion. Assuming that all the other parameters remained unchanged, the calculations were repeated, and an estimated emissions credits of 356 $\mathrm{MtCO_2e}$ for the UK was obtained. One explanation for this apparent contradiction is that reducing emissions, by reducing production and without decreasing consumption behaviors, would not reduce the global emissions, and would mostly result in the shifting of production activities to other regions which may have worse emissions intensities. A more detailed analysis could be performed in future work using an MRIO model.

The concepts of emission debt (EMDBT) and emission credits (EMCRD) allow us to define a percentage of non-greenness for each country, which we call the RED percentage. To do so, we use the Red emission scenario, which is based on the A1B AIM scenario, in order to define the projection interval, based on the assumption that the Green

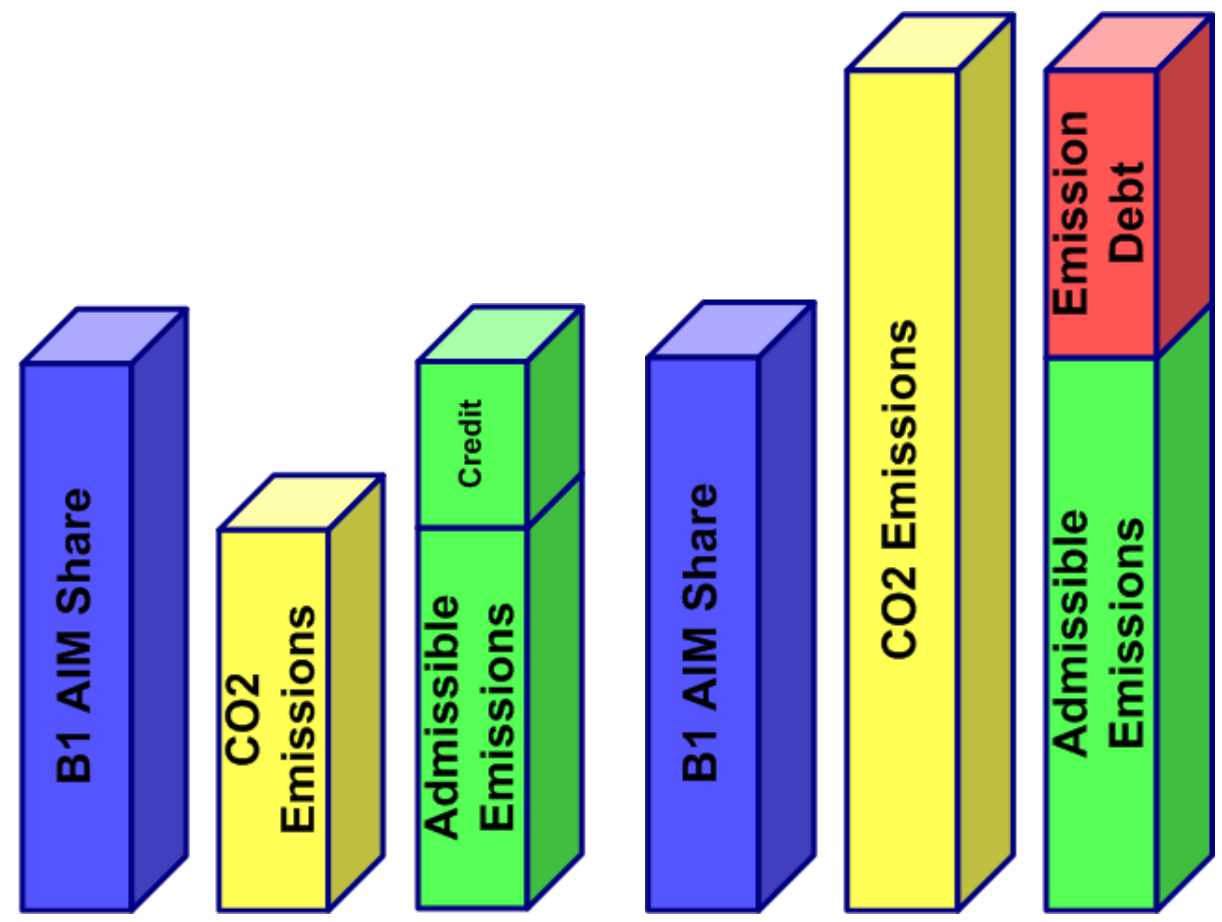

(a) (b)

Figure 8: An illustration of the emission credits and emission debt concepts: a) a country with positive emission credits. b) a country with positive emission debt.

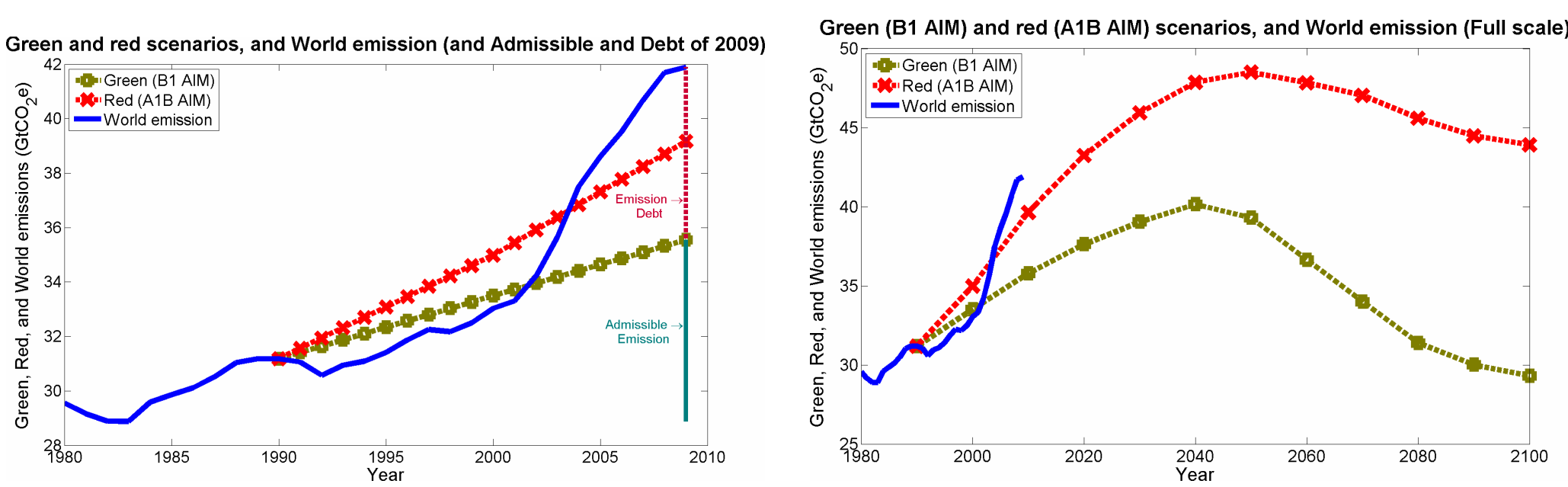

(a) (b)

Figure 9: a) An illustration of the admissible emissions and the emission debt concepts. b) The long-term trends of our Green and Red emissions scenarios (based on the B1 AIM and A1B AIM scenarios).

| Rank | Country | Admissible Emissions (MtCO2e) | Admissible Emissions per Capita (tCO2e) |
|---|---|---|---|
| **1** | **China** | **5,958** | **5.21** |
| **2** | **United States** | **5,671** | **22.68** |
| **3** | **India** | **2,801** | **3.25** |
| **4** | **Japan** | **1,657** | **13.42** |
| **5** | **Germany** | **1,256** | **15.92** |

(a)

| Rank | Country | Admissible Emissions per Capita (tCO2e) | Admissible Emissions (MtCO2e) |
|---|---|---|---|
| 1 | Zimbabwe | 0.59 | 6 |
| 2 | Congo, DRC | 0.94 | 38.8 |
| 3 | Burundi | 1.08 | 5.9 |
| 4 | Liberia | 1.11 | 3.4 |
| 5 | Mozambique | 1.26 | 17.8 |

(b)

Table 7: a) The 5 countries with largest amount of admissible emissions. b) The 5 countries with smallest amount of admissible emissions per capita.

emission scenario (B1 AIM) represents the 0% level, and the Red emission scenario (A1B AIM) the 100% level. We calculate the RED percentage as follows: first, we define the global emission debt margin as the difference in the amount of emissions of the A1B AIM and B1 AIM scenario; then, we calculate the emission debt margin of a country

| Rank | Country | Emissions Debt (MtCO2e) |
|---|---|---|
| **1** | **China** | **4,105** |
| **2** | **Russia** | **1,183** |
| **3** | **United States** | **909.4** |
| **4** | **Brazil** | **345.1** |
| 5 | Saudi Arabia | 307.3 |

(a)

| Rank | Country | Emissions Credit (MtCO2e) |
|---|---|---|
| **1** | **Japan** | **418.8** |
| 2 | France | 377.8 |
| **3** | **India** | **368.7** |
| **4** | **Germany** | **315.7** |
| 5 | United Kingdom | 295.2 |

(b)

| Rank | Country | Emissions Credit (MtCO2e) |
|---|---|---|
| **1** | **Japan** | **1,776** |
| **2** | **United States** | **1,426** |
| 3 | France | 821.4 |
| 4 | Italy | 778.9 |
| 5 | United Kingdom | 487.1 |

(c)

Table 8: a) The 5 countries with highest emission debt in 2009. b) Top 5 countries with highest emission credits in 2009. c) Top 5 countries with highest emission credits in 1990.

---

as a linear fraction of the admissible emissions:

$$\mathrm{EMDBT}^{\mathrm{MARG}}_{i,y} = \frac{\mathrm{ADMEM}_{i,y}}{\mathrm{GB1}_y}\left(\mathrm{RA1B}_y - \mathrm{GB1}_y\right) \tag{12}$$

where $\mathrm{EMDBT}^{\mathrm{MARG}}_{i,y}$ is the emission debt margin of country $i$ in year $y$. From this, the RED percentage can be easily defined as:

$$\mathrm{RED}_{i,y} = 100\frac{\mathrm{EMDBT}_{i,y}}{\mathrm{EMDBT}^{\mathrm{MARG}}_{i,y}} \tag{13}$$

where $\mathrm{RED}_{i,y}$ is an integer value which represents the RED percentage of country $i$ in year $y$. Automatically, the RED percentage will be 0% for a county without any emission debt. The distribution of the RED percentage over the world is shown in Figure 10. In order to make the variations over the map more visible, this distribution has been redrawn in Figure 10(b), considering South Africa's RED percentage as the maximum value. The countries with a 0% RED percentage are shown in gray (also shown in gray are countries that do not have a RED percentage because they have no IHDI index).

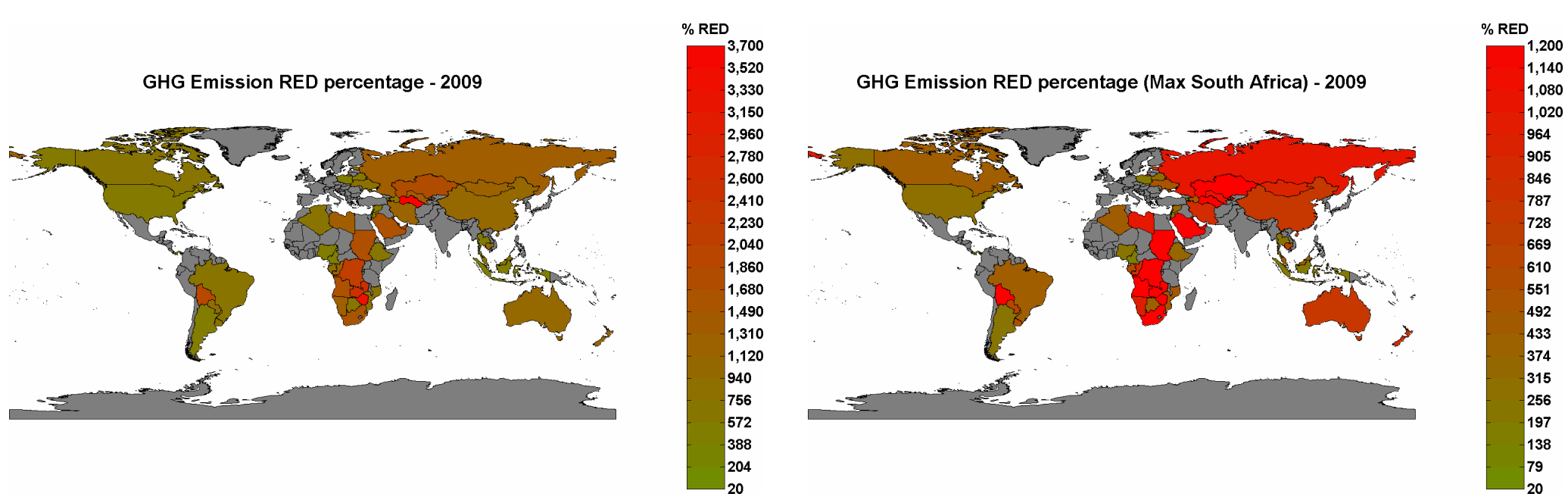

(a) (b)

Figure 10: a) The RED percentages distribution over the world in 2009. b) The same as in (a), but with South Africa's RED percentage as a maximum for better visualization. The percentages are divided by a factor of 100.

---

The top 5 countries in terms of the RED percentage are listed in Table 9(a). It is worth noting that all these top-5 countries, except Bahrain which is extremely rich, are extremely poor. This percentage shows by how much a country is over-emitting relative to its emission debt margin. For example, if a country has a total emissions of 36.30 $MtCO_2e$, and its admissible emission share is 25.50 $MtCO_2e$ (from the GB1 scenario), it will have an emission debt of 10.80 $MtCO_2e$. If its emission debt margin is 2.6 $MtCO_2e$ (from the difference between RA1B and GB1 scenarios), its RED percentage would be 10.80/2.6=416%. The allowed-emission share does not directly play in the calculations of the RED percentage. Implicitly, we can see that there is a conversion factor of 25.50/2.6 in this example, i.e, 100% of the over emissions is equal to 2.6/25.5=10.2% of the admissible emissions. This conversion factor is actually equal to the ratio of the difference between the green and red scenarios to the green scenario emissions at each year. In the next section, we use the RED percentage as the reference for imposing a carbon border tax. In the COI, South Africa has the highest RED percentage of $1,235\%$ in 2009 as can be seen from Table 9(b).

The trend of the B1 AIM share and the A1B AIM share, and the emissions of four countries are shown in Figure 11. China is an example of a country that has exceed both Green and Red emission shares. Indonesia is an example of a country that is within its Red (A1B AIM) emission share (65.1 $MtCO_2e$ more its Green emission share; its Red emission share is 76.1 $MtCO_2e$ more its Green emission share), but has exceeded its Green (B1 AIM) share (747.8 $MtCO_2e$). At the other end of the scale, we have Japan, whose emissions are below its Green (B1 AIM) share. The RED percentage for Japan was 0% in 2009 and all previous years. The last example is the United States. The impact there of the economic crisis not only on the emissions, but also on its Green and Red shares can easily be seen in Figure 11(d).

| **Rank** | **Country** | **RED %** |
|---|---|---|
| 1 | Turkmenistan | 3,608 |
| 2 | Zimbabwe | 3,230 |
| 3 | Bahrain | 2,274 |
| 4 | Zambia | 2,262 |
| 5 | Congo, DRC | 1,896 |

(a)

| Country | **RED %** | Country | **RED %** |
|---|---|---|---|
| **Australia** | **676** | **Indonesia** | **56** |
| **Brazil** | **326** | **Japan** | **0** |
| **Canada** | **322** | **Russia** | **986** |
| **China** | **676** | **South Africa** | **1,235** |
| **Germany** | **0** | **Switzerland** | **0** |
| **India** | **0** | **United States** | **157** |

(b)

Table 9: The RED percentage in 2009: a) for the top 5 countries, b) for the COI.

## 5. Carbon Tax and Emissions Trading

In this section, we introduce a Border Carbon Tax (BCT). Because of the highly competitive nature of the global market, we use a weak version of the RED percentage as the carbon tax. To do this, we define the BCT on the goods of country $i$ as:

$$\mathrm{BCT}_{i,y} = \mathrm{RED}_{i,y}/\mathrm{RED}_{\mathrm{BCT}} \tag{14}$$

where $\mathrm{BCT}_{i,y}$ is the border carbon tax percentage on the goods of country $i$ in year $y$, and $\mathrm{RED}_{\mathrm{BCT}}$ is the weak conversion factor from the RED to the BCT. We chose a value of 100 for $\mathrm{RED}_{\mathrm{BCT}}$ in this work to keep the BCT under the global annual growth. For example, for the United States, based on an $\mathrm{RED}_{\text{United States},2009} = 157\%$, the BCT on American goods would have been 1.6% in 2010. It is interesting to note that, among the COI, the United States has the second lowest BCT (excluding countries with a zero BCT). This means that it can impose a BCT on the goods of any other country whose BCT is higher than their own, even before establishing an equivalent internal carbon tax. It could, for example, have imposed a BCT of $6.8\% - 1.6\% = 5.2\%$ on Chinese goods in 2010, which was the difference between the BCTs of the two countries in 2009. In the following section, the impact of this tax on world emissions is analyzed by comparing two scenarios.

### 5.1. Carbon tax on imports

In this section, two scenarios are compared to evaluate the impact of the proposed tax system.

In both scenarios, it is assumed that the annual IHDIGDP growth of a country is linear, with an annual growth calculated on a period of 10 years between 2000 and 2009. For example, the annual IHDIGDP growth of China and the United States is 4.35% and $-2.49\%$ respectively. The maximum annual IDHIGDP growth over the same period corresponds to that of Qatar, with an annual IDHIGDP growth of 8.41%. For countries with lower IHDIGDP

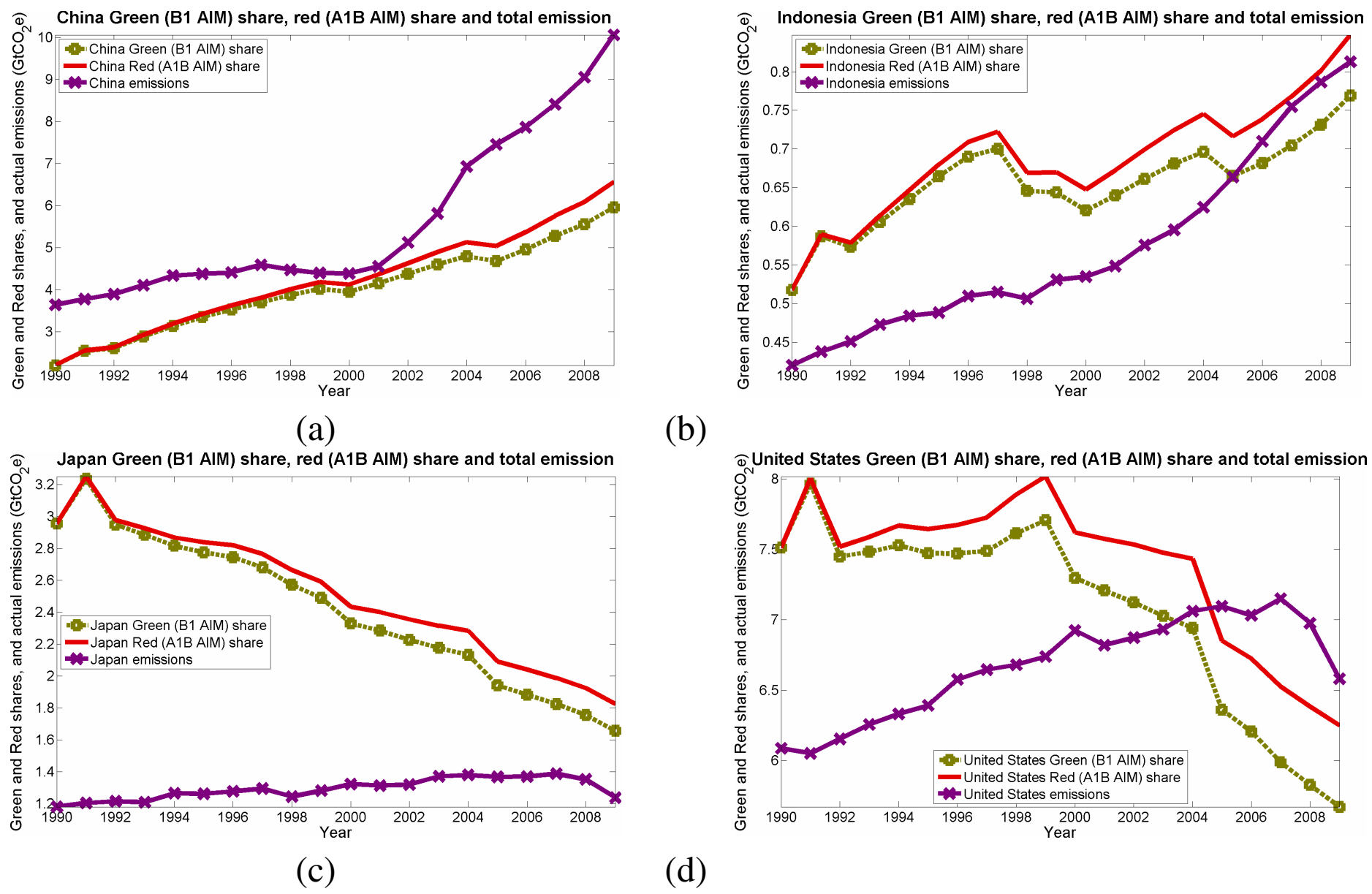

Figure 11: The trend of the Green (B1 AIM) share, the Red (A1B AIM) share, and the emissions of: a) China. b) Indonesia. c) Japan. d) the United States.

---

growth than the average global IHDIGDP annual growth from 1980 to 2009 (which is 1.67%), it is that average global IDHIGDP annual growth that is used in the calculations.

The two scenarios that are compared are the following:

- **CT scenario**: Imposition of the proposed tax, as in Equation (14). In this scenario, it is assumed that imposing this tax will decrease IHDIGDP growth, and at the same time reduce MGHG-INT. It is also assumed that for each 1% of tax imposed in a year, the IDHIGDP will decrease by 0.5%, and the MGHG-INT improves by 0.5% in the next year. It is assumed as well that technological breakthroughs and use of renewable energy will independently improve the MGHG-INT by 1.1% per year. Therefore, the governing equations of the scenario can be written as follows:

$$\left\{\begin{array}{rcl} \mathrm{IHDIGDP}_{i,y} & = & (1+\mathrm{IHDIGDP}'_{i,y-1})\times \\ & & (1-0.5\ \mathrm{BCT}_{i,y-1}/100)\times \\ & & \mathrm{IHDIGDP}_{i,y-1} \\ \mathrm{MGHGINT}_{i,y} & = & (1-0.5\ \mathrm{BCT}_{i,y-1}/100-0.011)\times \\ & & \mathrm{MGHGINT}_{i,y-1} \\ \mathrm{EM}_{i,y} & = & \mathrm{IHDIGDP}_{i,y}\ \mathrm{MGHGINT}_{i,y} \\ \mathrm{BCT}_{i,y} & = & \mathrm{RED}_{i,y}/\mathrm{RED}_{\mathrm{BCT}} \end{array}\right. \tag{15}$$

  where $\mathrm{IHDIGDP}'_{i,y}$ is the annual growth of $\mathrm{IHDIGDP}_{i,y}$, and $\mathrm{RED}_{i,y}$ is calculated based on Equations (9)-(13) in section 4.

- **NC scenario**: No BCT is imposed. However, here too, it is assumed that research and new technologies will improve the MGHG-INT by 1.1% per year.

$$\left\{\begin{array}{rcl} \mathrm{IHDIGDP}_{i,y} & = & (1+\mathrm{IHDIGDP}'_{i,y-1})\mathrm{IHDIGDP}_{i,y-1} \\ \mathrm{MGHGINT}_{i,y} & = & (1-0.011)\mathrm{MGHGINT}_{i,y-1} \\ \mathrm{EM}_{i,y} & = & \mathrm{IHDIGDP}_{i,y}\ \mathrm{MGHGINT}_{i,y} \\ \mathrm{BCT}_{i,y} & = & \mathrm{RED}_{i,y}/\mathrm{RED}_{\mathrm{BCT}} \end{array}\right. \tag{16}$$

The governing equations are solved over a 10-year period, until 2020, for both scenarios.

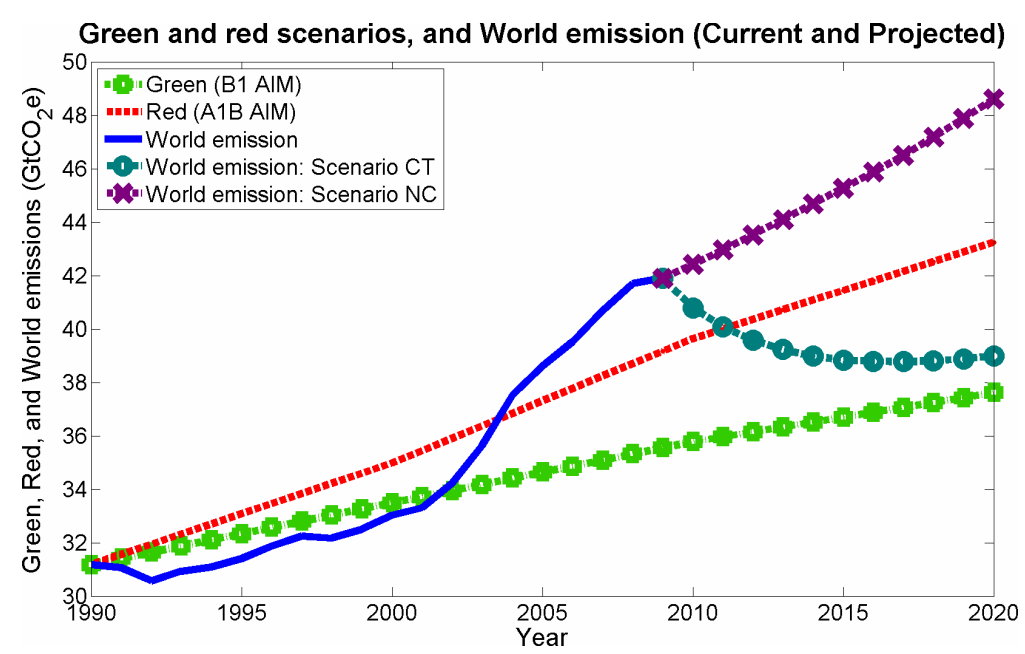

(a)

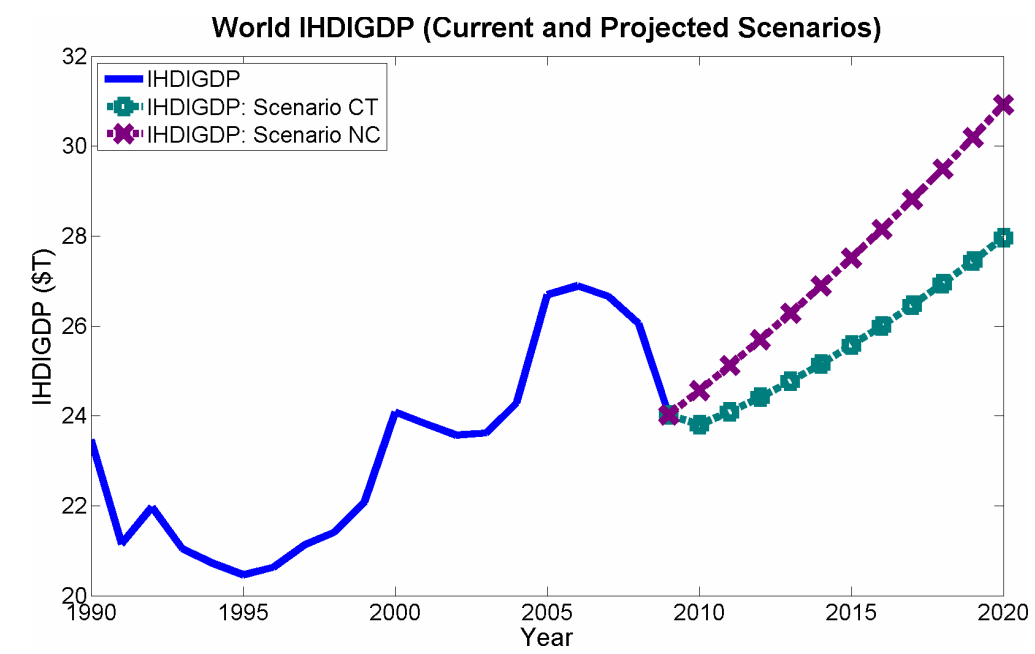

(b)

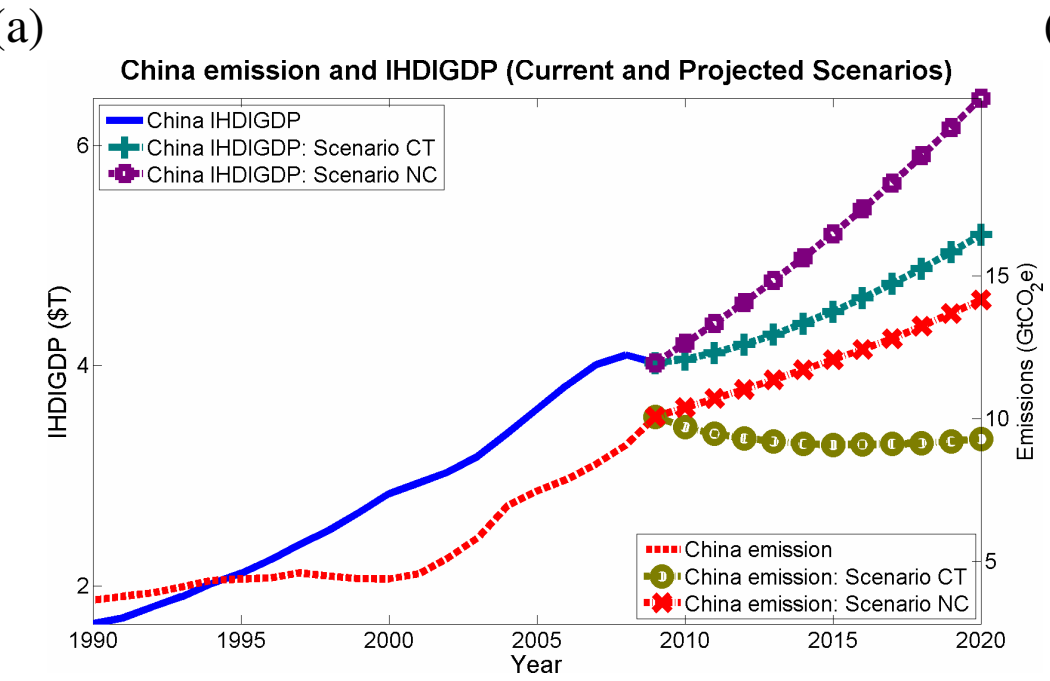

(c)

Figure 12: a) The impact of the proposed BCT on global emissions in the short term. In the CT scenario, the tax is implemented, and in the NC scenario, it is business as usual. b) The impact on global economic growth. c) The impact on China's economy and emissions.

---

In Figure 12(a), the difference between the two scenarios worldwide is depicted. As can be seen, by applying the proposed BCT, the emissions decrease globally, and even fall below the emissions in the Red (A1B AIM) emissions scenario. At the same time, the global economy will be growing, as shown in Figure 12(b). Although the NC scenario shows higher growth in this figure, it is just a projected result and does not take into account the risk of a global downturn resulting from climate change. The impact of the tax on rapidly growing China is presented in Figure 12(c) which shows the emissions and IHDIGDP of China in both scenarios. Again, in the CT scenario, China benefits from high economic growth, while its level of emissions is substantially lower.

For a better understanding of the impact of the BCT, the taxes of the COI at the beginning and end of the CT scenario are compared in 10. There is a definite decrease in all the taxes. It is worth noting that, in spite of global convergence toward the B1 AIM limit, they are still non-zero, because of the large proportion of these countries that are in global emission debt. In future work, we will confirm the results of these scenarios using multi-region input-output (MRIO) models.

### *5.2. Global Emissions Trading System*

In Table 11, the overall status of the world with respect to emissions, emission credits, and emission debt in 2009 is provided. The total unrecoverable emission debt is the difference between the total emission debt and the total emission credits. As can be seen from the table, there is a very considerable amount of unrecoverable emission debt (around 5, 793 $MtCO_2e$ in 2009), which means that there is a great deal of demand in the proposed emissions trading system, which will in turn lead to higher prices for emission credits. This will encourage countries to increase their emission credits not only to reduce their carbon tax, but also to sell it in the global ETS. Therefore, an emission trading

| COI | tax % |
|---|---|
| **Australia** | **6.8** |
| **Brazil** | **3.3** |
| **Canada** | **3.2** |
| **China** | **6.8** |
| **Germany** | **0** |
| **India** | **0** |
| **Indonesia** | **0.6** |
| **Japan** | **0** |
| **Russia** | **9.9** |
| **South Africa** | **12.4** |
| **Switzerland** | **0** |
| **United States** | **1.6** |

(a)

| COI | tax % |
|---|---|
| **Australia** | **2.2** |
| **Brazil** | **1.1** |
| **Canada** | **1.1** |
| **China** | **2.2** |
| **Germany** | **0** |
| **India** | **0** |
| **Indonesia** | **0** |
| **Japan** | **0** |
| **Russia** | **3** |
| **South Africa** | **3.5** |
| **Switzerland** | **0** |
| **United States** | **0.6** |

(b)

Table 10: The proposed border carbon tax for the COI: a) at the beginning of the CT scenario (2010), b) at the end of the CT scenario (2020).

| | Emission (MtCO2e) |
|---|---|
| World Emissions | 41,890 |
| Green (B1 AIM) Limit | 35,550 |
| Red (A1B AIM) Limit | 39,170 |
| Total Allowed Emissions | 35,550 |
| Total Emission Debt | 9,510 |
| Total Emission Credit | 3,717 |
| Total Unrecoverable Emission Debt | 5,793 |

Table 11: An overview of global emissions in 2009.

system at the international level for emission credits and emission debt will be very active. Countries with high debt will buy credits to reduce their RED percentage, and consequently the carbon tax on their goods.

We propose that land-use emissions should be first cleared in the ETS. In other words, a country with a debt in land-use emissions cannot buy emission credits for its emission debt before clearing its land-use emission debt. In this way, land-use emissions are neither promoted nor ignored, and it is left to individual countries to choose between land use and industrial development. It seems that a cumulative trading mechanism is required to maintain stability in the system and keep track of emissions.

### 5.3. *Carbon tax within a country*

A similar model as Equation (14) can be designed as a carbon tax system within a country. Such a system could be modeled on provincial or corporate slices. Also, it is suggested injecting the revenue from the BCT in each country into the growing of green initiatives. However, we recommend targeting the revenues in each sector to green industries in the same sector. However, the detailed discussion is beyond the scope of this work.

In terms of uncertainty in the data used, it has been observed that emissions at the international level are associated with a very low level of uncertainty [55, 1]. There are two possible reasons for this. One is the fact that a nation's emissions represent the sum of many small contributions from various sectors that may be unrelated. This lack of correlation can result in the cancellation of uncertainties. The second is that the total consumption or production of a sector can more easily be calculated than the composition of that consumption.

## 6. Conclusion and future prospects

A modified GHG intensity (MGHG-INT) indicator at the international level is introduced in this work. This measure, which is based on an IHDI-adjusted Gross Domestic Product (IDHIGDP) indicator, can be used to compare the emissions footprint of all nations, regardless of their place in the world economy. The IDHIGDP is obtained by combining the GDP and IHDI scores of countries, in order to account for all their activities. Using this universal measure, the non-greenness of the goods of each country is calculated in the form of its RED percentage. Then, a border carbon tax (BCT), along with a global ETS, are proposed. The impact of the proposed tax is analyzed over a period of one decade with promising results, in terms of the reduction in emissions, while preserving reasonable global growth. Land-use emissions are directed to the ETS in order to avoid any disagreement on the historical land-use emissions.

In future work, we will confirm the results of the proposed BCT scenario using a multi-region input-output (MRIO) model. The impact of the variable $RED_{BCT}$ conversion factor will be studied. Also, we think that the IHDI scores reported for some counties are higher than they should be, and so linearization of these scores in order to better reflect the differences in countries' development will be considered.

Furthermore, as our universal indicator targets a country as a whole, not its products or companies, disclosure of the carbon materiality of individual companies is not required. This is because, while national inventories are susceptible to error and manipulation, a nation's carbon materiality is expected to be reliable and robust. The only loophole might involve imported or indirect emissions. These emissions which will be addressed in another report using a Multi-Region Input/Output (MRIO) model. Including these emissions will help encourage countries with a high volume of imports to purchase low-emission goods.

Many special cases and issues should be addressed by a practical framework prior to implementation. For example, in terms of the Least Developed Countries (LDCs), some exceptions can easily be added to the border taxes or tax adjustments. Moreover, thanks to the unilateral nature of the proposed solution, any other country can voluntarily waive the taxes levied on the LDCs. However, we do not suggest any change or exception to the definition of the universal indicators, because they should constitute an indicator of development and reflect the true situation of all countries.

It is worth noting that market-based mechanisms, in particular border adjustments, have been discussed by policy-makers and academia for a long time as possible mechanisms for moving toward a sustainable world, including: the Durban Platform for Enhanced Action at the international level, the aviation carbon tax, carbon border cost leveling (CBCL), the carbon inclusion mechanism (CIM) at the EU level, and the U.S. carbon tax at a national level, which could eventually bear fruit. We hope that our work will constitute a modest contribution to the international effort toward creating a better world.

## Acknowledgments

The authors thank the NSERC of Canada for their financial support under grant CRDPJ 424371-11. The authors also thank the anonymous reviewers for their valuable and insightful comments and suggestions to improve the quality of the paper.

## A. Notations and Basic Concepts

### *A.1. $CO_2$ Emissions and GHG Emissions*

As mentioned in the introduction, we only consider energy-related $CO_2$ emissions in the carbon tax calculation. These data, which were retrieved from US Energy Information Administration databases[2], were gathered over the 30-year period between 1980 and 2009. They are restricted to emissions from fuel combustion. The world profile of $CO_2$ emissions is shown in Figure 13(a).

For the other GHGs, the data from the World Bank databases are used. In this study, we consider methane ($CH_4$) emissions from human activities such as agriculture, and from industrial methane production[3], $NO_X$ emissions from the burning of agricultural biomass, industrial activities, and livestock management[4], as well as HFC, PFC, and $SF_6$ (HPS) emissions[5] are considered. It is worth noting that these emissions are measured in equivalent gigatonnes of $CO_2$ ($GtCO_2e$). According to the IPCC's 2001 Third Assessment Report, the Global Warming Potential (GWP) for $CH_4$ is 23 for a time horizon of 100 years [51]. The GWP for $N_2O$ is estimated to be 296 for the same time horizon. The world profiles in terms of methane, $NO_X$, and HPS emissions in 2009 are shown in Figures 13(b), 13(c), and 13(d). The total GHG emission distribution across the globe in 2009 in $GtCO_2e$ is shown in Figure 2(a) in section 2.

### *A.2. Human Development Index (HDI) and Inequality-adjusted HDI (IHDI)*

As discussed in the section on motivations, we are looking for an independent indicator of national activity in order to modify countries' GDP. We have chosen an indicator based on the Human Development Index (HDI), which is an well-known independent indicator used in the United Nations Development Programme (UNDP) that summarizes human development status [42]. It measures the average achievements of a country in three basic dimensions of human development: a long and healthy life, access to knowledge, and a decent standard of living. The HDI can be seen as a "mean of means", where it first finds the average income achievement, the average educational achievement, and the average health achievement, and then takes the average of the three to obtain the HDI level [41]. The HDI is available from the UNDP and UNdata.

[2] http://www.eia.gov/cfapps/ipdbproject/IEDIndex3.cfm?tid=90&pid=44&aid=8
[3] http://data.worldbank.org/indicator/EN.ATM.METH.KT.CE/countries?page=1
[4] http://data.worldbank.org/indicator/EN.ATM.NOXE.KT.CE/countries
[5] http://data.worldbank.org/indicator/EN.ATM.GHGO.KT.CE/countries?page=1

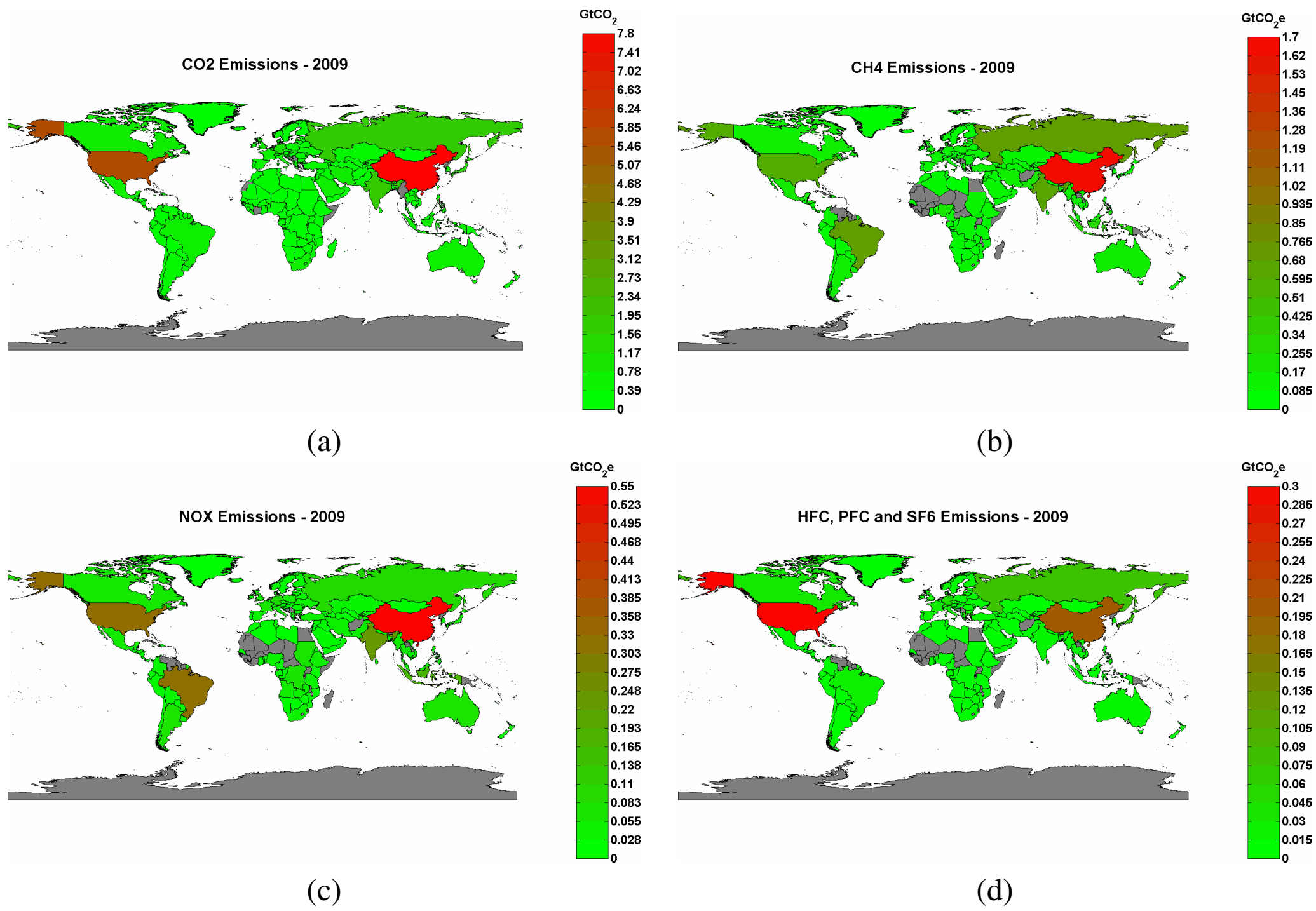

Figure 13: a) $CO_2$ emissions from the energy consumption in 2009. b) $CH_4$ emissions in 2009. c) $NO_X$ emissions in 2009. d) HFC, PFC and SF6 emissions in 2009.

Although the HDI is a promising indicator and considers variations in terms of whether or not goods and services can be bought, it ignores inequalities within a nation. It is reasonable to assume that the level of development is lower in countries with a higher level of inequality. The Inequality-adjusted HDI (IHDI) is an attempt to address this drawback of HDI [41, 42][6], and we use it in our calculations as an indicator of national development per capita.

To quantify the total activity level of a nation, we introduce an indicator called HDIxCapita, which is defined as the nation's IHDI multiplied by the its population snapshot. The two activity indicators, GDP (PPP) and HDIxCapita, are combined in a new indicator of activity in section 3, which we call the IHDI-adjusted Gross Domestic Product (IHDIGDP).

*A.3. B1 AIM Scenario*

As discussed in section 4, we needed two emissions scenarios, which we labeled Green and Red, in order to calculate the RED percentage of nations. Although the choice of these scenarios is arbitrary, we elected to build them based on the IPCC emissions scenarios. The IPCC integrates their B1 environmentally friendly scenarios [44, 45, 46, 47, 48, 49, 50, 51, 52, 53, 54] at the global level. Although they reflect rapid economic growth, these scenarios anticipate equally rapid changes toward a service and information economy. In this way, dematerialization and the introduction of clean technologies will help the world to maintain its economic growth, while giving the environment an opportunity to recover. These B1 scenarios assume that global solutions will be developed for economic, social and environmental stability with improving equity. They see the world's population rising to 9 billion in 2050 and then declining [53]. The same trend is assumed for GHG emissions. Among several models they use, the Asian-Pacific Integrated Model (AIM) [43] is a large-scale computer simulation for scenario analysis of GHG emissions and the impacts of global warming designed for the Asia-Pacific region. This model is linked to a world model, so that global estimates can be made. AIM comprises three main models: the GHG emissions model, the global climate change model, and the climate change impact model [43].

[6] `http://hdrstats.undp.org/en/indicators/73206.html`

The Green scenario is our environmentally friendly scenario, for which we used the IPCC's B1 AIM scenario data[7]. We distributed the $CO_2$ emissions accumulated between 1990 and 2100 evenly over the years. In the B1 AIM scenario, the $CH_4$ emissions peak at 449 Mega tonnes of methane in 2050, which is equivalent to $23 \times 449 = 10327$ $MtCO_2e$, and to $100 \times (10327/33270) = 31.04\%$ of $CO_2$ emissions. Similar trends exist for other GHG emissions. Therefore, to define the Green scenario for total GHG emissions, we multiplied the B1 AIM values by 2. Finally, we added the world emissions of 1990 to arrive at an absolute value for Green scenario emissions:

$$\mathrm{GB1}_y = \mathrm{EM}_{\mathrm{world},1990} + 2\left(\mathrm{B1}_y^{\mathrm{AIM}} - \mathrm{B1}_{1990}^{\mathrm{AIM}}\right) \tag{17}$$

where $\mathrm{GB1}_y$ is the GHG emissions limit of the Green scenario, $\mathrm{EM}_{\mathrm{world},1990}$ is the world GHG emissions in 1990, and $\mathrm{B1}_y^{\mathrm{AIM}}$ is the B1 AIM scenario $CO_2$ emissions in year $y$.

For the Red scenario, we used the same procedure, but with A1B AIM scenario values. The A1B scenarios are a subset of the A1 scenarios that focus more on the balance between fossil fuels and other energy sources. The A1 scenarios, like the B1 scenarios, assume very rapid economic growth and a global population that peaks at mid-century and declines thereafter, along with the equally rapid introduction of new and more efficient technologies. Major underlying themes are convergence among regions, capacity building, and increased cultural and social interaction, with a substantial reduction in regional differences in per capita income. The main difference between the A1 and B1 scenarios is the de-materialization aspect of the B1 scenarios, which see significant changes in economic structures, leading to a service and information economy, with consequent reductions in material-dependent production, and the introduction of clean and resource-efficient technologies [53]. We have built our Red scenario based on the A1B AIM scenario:

$$\mathrm{RA1B}_y = \mathrm{EM}_{\mathrm{world},1990} + 2\left(\mathrm{A1B}_y^{\mathrm{AIM}} - \mathrm{A1B}_{1990}^{\mathrm{AIM}}\right) \tag{18}$$

where $\mathrm{RA1B}_y$ is the GHG emissions limit of the Red scenario and $\mathrm{A1B}_y^{\mathrm{AIM}}$ is the A1B AIM scenario $CO_2$ emissions in year $y$. The Green and Red scenarios are used in section 4 to calculate the RED percentage.

In all the tables, the countries of interest (COI) are highlighted. The COI is composed of countries with global or regional influence, and includes Australia, Brazil, China, Canada, Germany, India, Indonesia, Japan, Russia, South Africa, the United States, and Switzerland.

## B. Motivations

Here, we discuss the global threats which motivate for a new universal measure in details.

### *B.1. Threats: Global disagreement on emission reduction goals, measures, and procedures.*

In a highly interconnected world economy, global goals and measures can result in advantages for exporting countries. This will force, and has forced, many developed countries to abandon such GHG reduction goals and measures, and adopt the strategy that best suits their needs. Canada, for example, changed its goal from meeting the Kyoto targets to the new goal of reducing emissions intensity in 2006 [56], and officially withdraw from the protocol in December 2011[8] [57]. In a competitive global economy, these individually set goals prevent countries from reaching the global agreement that is the key to future emissions reductions.

The implementation of a universal measure for all countries has not occurred, mainly because of two ongoing debates: 1) between developing countries and developed countries on historical responsibility, in terms of the cumulative emissions and land use of the developed countries; and 2) on the transfer of the technology required to improve production efficiency and to reduce the emissions of developing countries. Our proposed measure, MGHG-INT, provides a unified view of all the countries involved, and eliminates any debate on the differences between them:

- Production vs. consumption: It seems that the major difference between developing countries and developed countries is in their production and consumption habits. The proposed MGHG-INT uses the term activity instead, and so covers both production and consumption in a universal way.
- Historical (cumulative) emissions levels: The issue of the cumulative footprint of countries is addressed and discussed in section C of the appendix. In brief, it has been observed that, on a scale of two decades, the cumulative impact is, in fact, less than 1 percent. We will therefore ignore it in the main part of this work.
- Historical emissions (land use): The proposed measure does not consider land-use emissions (and de-emissions). We suggest that these emissions be submitted to the proposed global emissions trading system (ETS), as discussed in section 5.2. The developing countries are allowed to produce land-use emissions. However, they cannot clear their GHG emissions before clearing their land use emissions.
- Technology transfer: it seems that technology transfer, in the form of adapting best practices, could boost the movement toward a greener earth if a suitable control mechanism is implemented. This is beyond the scope of this work, but will be considered in the future by proposing technological leapfrogging for developing countries [58, 59] while respecting intellectual property rights.

[7] `http://sres.ciesin.org/final_data.html`
[8] `http://www.cbc.ca/news/interactives/canada-kyoto/`

### *B.2. Threats: De-industrialization of Europe*

While there has been much debate and discussion on whether or not the European emissions reduction policy has accelerated Europe's de-industrialization [9], it is obvious that a high volume of manufactured goods is flowing from the BRICS countries (Brazil, the Russian Federation, India, China, and South Africa) to Europe [60]. In particular, some industries are facing carbon leakage through so-called competitiveness channel [22]. For example, a high rate of leakage has been detected in the steel and cement industries [61, 62]. In terms of European de-industrialization, many parameters, such as fewer work hours and high unemployment rates (the effect of an aging population), have been mentioned as key factors [60]. However, the advantages of low wages and low prices of the BRICS countries, as well as the use of unpenalized non-green energy sources, may also be factors [60, 63, 64]. The de-industrialization of Europe, if it exists, may accelerate the increase in global GHG emissions, because it moves the industries it loses to those regions of the world that have very little or no environmental supervision. As a result, in a competitive global economy, more non-green energy will be generated from low-cost, non-green sources.

At the same time, it has been argued that climate change discussions are overly focused on its environmental impacts, and ignore its socio-economic consequences [63]. As mentioned in the previous subsection, a uniform and universal emission control policy can address this deficiency, and enable all global players to pursue their economical growth in a sustainable way. We hope that the MGHG-INT and our proposed Border Carbon Tax (BCT) and ETS can form the basis for such a global emission control policy.

### *B.3. Threats: Leakage and hidden emissions in exports/imports and transport*

Although a number of European countries have been able to work toward their commitments on emission reduction within the Kyoto protocol, it has been observed by many researchers that some of this reduction has been accomplished by outsourcing the emissions to other countries [9]. For example, in [65], using a UK-specific multi-region input-output (MRIO) model, it has been observed that the net $CO_2$ emissions embedded in UK imports increased from 4.3% of producer emissions in 1992 to a maximum of 20% in 2002. The total estimated UK carbon footprint in 2004 was 730 Mega tonnes of carbon dioxide ($MtCO_2$) and 934 Mega tonnes of carbon dioxide equivalent ($MtCO_2e$) for all GHGs [65]. It has also been observed that one-third of China's emissions are related to goods to be exported [13], and globally 23% of $CO_2$ emissions from the burning of fossil fuels is used in the production of goods that were consumed in a different country [66]. This suggests that consumption-based emission analysis should be performed, in order to see and control hidden carbon leakage [66]. However, as mentioned above, consumption-based analysis cannot cover all countries, and so we argue that analysis should be based on activities, as we explain in section 3.

### B.4. Threats: Population

Population growth alone is a threat to the earth's sustainability. Any measure that promotes an increase in the world's population should be avoided, and so, because of population growth, the measure of GHG emissions per capita is a dangerous one. Instead, in this work, a measure of the activity of a population based on its level of development is used. In our calculations, we use a snapshot of the population in a reference year (1990 in our case). Among a number of global human development measures, the Inequality-adjusted Human Development Index (IHDI) is well-known and accepted (see section A.2 of the appendix for details). We multiply the IHDI by the population snapshot, and use that figure in the calculation, instead of the current population (see section 3).

### B.5. Threats: Administration cost

One of parameters that is used to avoid the adoption of an emission control policy is administrative cost. Unlike a traditional border carbon tax, whose associated administrative costs could be as high as a proportion of price of goods [21], our solution, which is independent of any particular product, targets a country as a whole. Now, a country's bureaucracy is responsible for implementing mechanisms to reduce their total emissions (perhaps implementing a small scale of our solution within their country, as will be discussed in subsection 5.3). If it fails to do so, the country will face the migration of industries and markets to other regions of the world with more efficient infrastructure and better emission reduction policies.

A firm could argue that it is producing a small-footprint product in a country with a large footprint, and therefore the carbon tax calculated for this country is unfair to the firm. However, it should be taken into account that this firm has a hidden footprint associated with its employees. In a firm with 10,000 employees, there could be 50,000 or so residents, including extended family members, living a non-green lifestyle in a country with an inefficient infrastructure. At the same time, it is worth noting that the average footprint of an employee in a firm in a developing country should be considered to be several times larger than the footprint per capita of that country, because of the high degree of social inequality in these countries.

## C. MGHG-CINT: Modified GHG Emission Cumulative Intensity

Although we are at the beginning of the global industrial boom of the 21$^{st}$ century, it can be argued that the cumulative emissions of countries should be considered rather than their current emission levels. This argument makes sense, as the GHG lifetimes in the atmosphere are usually long; for example, that of $CH_4$ is estimated to be a decade. To take this point into consideration, we have re-calculated all the variables in this section, to express them in cumulative form. The calculation period we selected is from 1990 to 2009. As the IHDIGDP is a normalized figure, no conversion on the IHDIGDP values has been applied. The GHG emissions and the Green and Red scenarios over the calculation period have been summed to obtain their cumulative versions. Table 12 presents the top 5 countries and the COI countries in terms of their RED percentages. As can be concluded from the table, the difference between the annual (Table 9) and cumulative RED percentages is small, and so the annual values are sufficient, and perhaps more rewarding, for developing a short-term policy. Also, in a future work, the authors will perform the cumulative study over larger periods of time (centuries) by extrapolating the IHDI data to the past.

| Rank | Country | REDC % |
|---|---|---|
| 1 | Congo, DRC | 6,712 |
| 2 | Turkmenistan | 6,101 |
| 3 | Zambia | 5,266 |
| 4 | Zimbabwe | 4,357 |
| 5 | Angola | 3,939 |

(a)

| COI | REDC % | COI | REDC % |
|---|---|---|---|
| **Australia** | **1,052** | **Indonesia** | **0** |
| **Brazil** | **0** | **Japan** | **0** |
| **Canada** | **450** | **Russia** | **2,128** |
| **China** | **798** | **South Africa** | **2,029** |
| **Germany** | **0** | **Switzerland** | **0** |
| **India** | **0** | **United States** | **143** |

(b)

Table 12: The RED percentage in 2009, cumulated from 1990: a) for the top 5 countries, b) for the COI.

Even if developed countries had historically produced no emissions, as a global courtesy, these countries should give developing countries some credits in an effort to redress the development balance, as this will prove to be a key element in global sustainability. However, there are some aspects of this issue which suggest that these credits should be highly supervised and controlled. First, although emissions have historically been produced by developed countries, the main outcome of development, which is knowledge, is shared (and should be shared) among all the nations of the world. Medical and technological advances, for example, have saved many lives everywhere. Estimating of this share will require a thorough analysis, and can shed some light on this debate. Second, even if a developing country is hypothetically entitled to some emission credits, then there should also be some sort of "additionality" constraint. Additionality is a parameter that is found in all emission quantification and reporting standards and protocols. In our case, the credits should not allow that country to use old, inefficient practices with a large environmental footprint. Third, to the commonly held view that a country is considered as a fundamental unit, or *atom*, we should add that it can be split into two major subatomic parts: i) the *Powerful* and ii) the *Rest*. Unconstrained or uncontrolled credits could be used by the *Powerful* without a significant benefit to the *Rest*, as has happened several times in recent decades in other international moves to help developing countries. This behavior is expected because of the high degree of corruption and irresponsibility that exists in some of these countries. As the main goal of the credit entitlement for developing countries is to improve the quality of life of the *Rest*, some controls or mechanisms should be built into any strategy designed to achieve this goal. This is why the authors considered the IHDI index for this purpose, which is an index of inequality, in the proposed MGHG-INT indicator. Fourth, the traditional clustering of countries into two classes, developed and developing, no longer seems to reflect reality, and now perhaps various indicators and indices should be included to create a unique picture of each country. And finally, in the result obtained, we can see that the extent of the increase in BCTs for developing countries is very high compared to that of the developed ones when a cumulative study is used. For example, China and the USA have a BCT of 6.8% and 1.6% respectively, based on their 2009 emissions. However, if we consider a cumulative study from 1990 to 2009, their BCTs would be 8% and 1.4% respectively. This could mainly be because of very low level of IHDI in the developing countries over the past.

# A Modified GHG Intensity Indicator: Toward a Sustainable Global Economy based on a Carbon Border Tax and Emissions Trading (Supplementary Material)

Reza Farrahi Moghaddam, Fereydoun Farrahi Moghaddam and Mohamed Cheriet

*Synchromedia Laboratory for Multimedia Communication in Telepresence,*
*École de technologie supérieure, Montreal (QC), Canada H3C 1K3*
*Tel.: +1(514)396-8972*
*Fax: +1(514)396-8595*
*rfarrahi@synchromedia.ca, imriss@ieee.org, ffarrahi@synchromedia.ca, mohamed.cheriet@etsmtl.ca*

**Supplementary Material: Full Tables**

| Country | GDP (PPP) | Country | GDP (PPP) | Country | GDP (PPP) | Country | GDP (PPP) | Country | GDP (PPP) |
|---|---|---|---|---|---|---|---|---|---|
| Afghanistan | 25.04 | Congo, DRC | 21.35 | **India** | **3,645** | Montenegro | 6.59 | Spain | 1,358 |
| Albania | 22.84 | Congo | 15.53 | **Indonesia** | **961.4** | Morocco | 145.4 | Sri Lanka | 96.67 |
| Algeria | 240.7 | Costa Rica | 48.65 | Iran | 802.7 | Mozambique | 20.19 | St. Kitts and Nevis | 0.688 |
| Angola | 104.6 | Ivory | 35.75 | Iraq | 111.4 | Myanmar | 71.96 | St. Lucia | 1.766 |
| Antigua and Barbuda | 1.472 | Croatia | 78.43 | Ireland | 172.5 | Namibia | 13.85 | St. Vincent and the Grenadines | 1.083 |
| Argentina | 582.9 | Cyprus | 22.73 | Israel | 207.8 | Nepal | 33.92 | Sudan | 94.28 |
| Armenia | 16.28 | Czech Republic | 253 | Italy | 1,734 | The Netherlands | 659 | Suriname | 4.469 |
| **Australia** | **850.6** | Denmark | 195.8 | Jamaica | 23.76 | New Zealand | 114.9 | Swaziland | 5.893 |
| Austria | 322.5 | Djibouti | 1.995 | **Japan** | **4,107** | Nicaragua | 16.79 | Sweden | 333 |
| Azerbaijan | 85.65 | Dominica | 0.744 | Jordan | 33.17 | Niger | 10.18 | **Switzerland** | **313.4** |
| The Bahamas | 8.793 | Dominican Republic | 80.21 | Kazakhstan | 181.8 | Nigeria | 345.4 | Syria | 103.1 |
| Bahrain | 28.28 | Ecuador | 110.4 | Kenya | 62.3 | Norway | 251.8 | Taiwan | 734.5 |
| Bangladesh | 241.6 | Egypt | 468.9 | Kiribati | 0.601 | Oman | 72.08 | Tajikistan | 13.71 |
| Barbados | 6.199 | El Salvador | 42.84 | South Korea | 1,362 | Pakistan | 439.4 | Tanzania | 54.36 |
| Belarus | 120.8 | Equatorial Guinea | 23.79 | Kosovo | 11.4 | Panama | 40.87 | Thailand | 539.2 |
| Belgium | 383.1 | Eritrea | 3.514 | Kuwait | 132.6 | Papua New Guinea | 13.83 | East Timor | 2.85 |
| Belize | 2.574 | Estonia | 23.72 | Kyrgyzstan | 12.07 | Paraguay | 28.62 | Togo | 5.723 |
| Benin | 13.53 | Ethiopia | 78.99 | Laos | 14.43 | Peru | 251 | Tonga | 0.742 |
| Bhutan | 3.596 | Fiji | 3.827 | Latvia | 32.32 | Philippines | 324.3 | Trinidad and Tobago | 25.85 |
| Bolivia | 45.52 | Finland | 178.6 | Lebanon | 54.71 | Poland | 688.2 | Tunisia | 95.52 |
| Bosnia and Herzegovina | 29.8 | France | 2,094 | Lesotho | 3.193 | Portugal | 241.3 | Turkey | 879.3 |
| Botswana | 26 | Gabon | 21.07 | Liberia | 1.593 | Qatar | 128.3 | Turkmenistan | 33.47 |
| **Brazil** | **2,002** | The Gambia | 3.273 | Libya | 86.13 | Romania | 255 | Tuvalu | 0.036 |
| Brunei | 19.39 | Georgia | 20.9 | Lithuania | 55.32 | **Russia** | **2,118** | Uganda | 39.7 |
| Bulgaria | 95.72 | **Germany** | **2,814** | Luxembourg | 39.37 | Rwanda | 11.31 | Ukraine | 290.1 |
| Burkina Faso | 18.72 | Ghana | 58.06 | Macedonia | 19.67 | Samoa | 1.045 | United Arab Emirates | 236.8 |
| Burundi | 3.241 | Greece | 330 | Madagascar | 19.61 | Sao Tome and Principe | 0.295 | United Kingdom | 2,126 |
| Cambodia | 28.2 | Grenada | 1.103 | Malawi | 12.06 | Saudi Arabia | 593.9 | **United States** | **14,120** |
| Cameroon | 42.63 | Guatemala | 67.72 | Malaysia | 383.1 | Senegal | 22.7 | Uruguay | 43.82 |
| **Canada** | **1,278** | Guinea | 10.5 | Maldives | 2.508 | Serbia | 77.97 | Uzbekistan | 78.37 |
| Cape Verde | 1.793 | Guinea-Bissau | 1.708 | Mali | 15.9 | Seychelles | 1.914 | Vanuatu | 1.102 |
| Central African Republic | 3.305 | Guyana | 5.141 | Malta | 9.945 | Sierra Leone | 4.455 | Venezuela | 348.6 |
| Chad | 16.37 | Haiti | 11.97 | Mauritania | 6.298 | Singapore | 252.6 | Vietnam | 256.5 |
| Chile | 242.7 | Honduras | 32.41 | Mauritius | 17.2 | Slovakia | 114.4 | Yemen | 58.14 |
| **China** | **9,057** | Hong Kong | 302.1 | Mexico | 1,471 | Slovenia | 55.38 | Zambia | 18.45 |
| Colombia | 413.4 | Hungary | 183.6 | Moldova | 10.18 | Solomon Islands | 1.527 | Zimbabwe | 4.959 |
| Comoros | 0.776 | Iceland | 12.13 | Mongolia | 10.28 | **South Africa** | **504.9** | | |

Table 1. The GDP (PPP) of nations (in B$ International) in 2009.

| Country | Population ($\times 10^6$) | Country | Population ($\times 10^6$) | Country | Population ($\times 10^6$) | Country | Population ($\times 10^6$) | Country | Population ($\times 10^6$) |
|---|---|---|---|---|---|---|---|---|---|
| Afghanistan | 22.8 | Congo, DRC | 41.26 | **India** | **862.2** | Montenegro | NaN | Spain | 38.84 |
| Albania | 3.197 | Congo | 2.232 | **Indonesia** | **179.8** | Morocco | 24.04 | Sri Lanka | 16.27 |
| Algeria | 25.02 | Costa Rica | 3.81 | Iran | 54.5 | Mozambique | 14.15 | St. Kitts and Nevis | 0.042 |
| Angola | 10.78 | Ivory | 11.72 | Iraq | 27.1 | Myanmar | 40.79 | St. Lucia | 0.134 |
| Antigua and Barbuda | 0.063 | Croatia | 4.381 | Ireland | 3.506 | Namibia | 1.345 | St. Vincent and the Grenadines | 0.106 |
| Argentina | 32.53 | Cyprus | 0.579 | Israel | 4.514 | Nepal | 19.11 | Sudan | 25.75 |
| Armenia | 3.211 | Czech Republic | 10.2 | Italy | 56.69 | The Netherlands | 14.95 | Suriname | 0.461 |
| **Australia** | **17.17** | Denmark | 5.135 | Jamaica | 2.373 | New Zealand | 3.411 | Swaziland | 0.898 |
| Austria | 7.678 | Djibouti | 0.507 | **Japan** | **123.4** | Nicaragua | 5.455 | Sweden | 8.568 |
| Azerbaijan | 7.459 | Dominica | 0.072 | Jordan | 3.468 | Niger | 7.731 | **Switzerland** | **6.712** |
| The Bahamas | 0.254 | Dominican Republic | 7.148 | Kazakhstan | 14.85 | Nigeria | 90.56 | Syria | 12.12 |
| Bahrain | 0.48 | Ecuador | 10.63 | Kenya | 24.14 | Norway | 4.25 | Taiwan | 20.4 |
| Bangladesh | 115.6 | Egypt | 51.36 | Kiribati | 0.072 | Oman | 1.63 | Tajikistan | 5.536 |
| Barbados | 0.25 | El Salvador | 5.11 | South Korea | 42.87 | Pakistan | 108.4 | Tanzania | 25.47 |
| Belarus | 9.48 | Equatorial Guinea | 0.452 | Kosovo | NaN | Panama | 2.394 | Thailand | 56.3 |
| Belgium | 9.977 | Eritrea | 3.192 | Kuwait | 2.13 | Papua New Guinea | 3.758 | East Timor | 0.782 |
| Belize | 0.189 | Estonia | 1.34 | Kyrgyzstan | 4.467 | Paraguay | 4.089 | Togo | 3.961 |
| Benin | 5.332 | Ethiopia | 48.29 | Laos | 4.207 | Peru | 21.75 | Tonga | 0.096 |
| Bhutan | 0.549 | Fiji | 0.724 | Latvia | 2.261 | Philippines | 61.5 | Trinidad and Tobago | NaN |
| Bolivia | 6.514 | Finland | 4.998 | Lebanon | 2.83 | Poland | 38.2 | Tunisia | 8.154 |
| Bosnia and Herzegovina | 3.654 | France | 56.71 | Lesotho | 1.723 | Portugal | 9.996 | Turkey | 52.44 |
| Botswana | 1.28 | Gabon | 0.93 | Liberia | 3.071 | Qatar | 0.422 | Turkmenistan | 3.861 |
| **Brazil** | **146.6** | The Gambia | 0.863 | Libya | 4.365 | Romania | 23.54 | Tuvalu | NaN |
| Brunei | 0.253 | Georgia | 4.315 | Lithuania | 3.34 | **Russia** | **147.7** | Uganda | 17.73 |
| Bulgaria | 8.718 | **Germany** | **78.89** | Luxembourg | 0.382 | Rwanda | 7.155 | Ukraine | 45.71 |
| Burkina Faso | 8.532 | Ghana | 14.31 | Macedonia | 1.922 | Samoa | NaN | United Arab Emirates | 1.844 |
| Burundi | 5.46 | Greece | 10.16 | Madagascar | 12.03 | Sao Tome and Principe | 0.115 | United Kingdom | 57.24 |
| Cambodia | 8.487 | Grenada | 0.096 | Malawi | 10.04 | Saudi Arabia | 15.19 | **United States** | **250** |
| Cameroon | 11.52 | Guatemala | 8.211 | Malaysia | 18.1 | Senegal | 7.977 | Uruguay | 3.094 |
| **Canada** | **27.63** | Guinea | 6.147 | Maldives | 0.212 | Serbia | 7.382 | Uzbekistan | 21.36 |
| Cape Verde | 0.345 | Guinea-Bissau | 0.98 | Mali | 7.987 | Seychelles | 0.07 | Vanuatu | 0.148 |
| Central African Republic | 2.945 | Guyana | 0.751 | Malta | 0.354 | Sierra Leone | 3.528 | Venezuela | 19.5 |
| Chad | 5.648 | Haiti | 7.076 | Mauritania | 1.959 | Singapore | 3.135 | Vietnam | 66.02 |
| Chile | 13.1 | Honduras | 4.901 | Mauritius | 1.059 | Slovakia | 5.308 | Yemen | 12.09 |
| **China** | **1,143** | Hong Kong | 5.752 | Mexico | 83.23 | Slovenia | 1.978 | Zambia | 7.91 |
| Colombia | 34.13 | Hungary | 10.38 | Moldova | 3.568 | Solomon Islands | 0.321 | Zimbabwe | 10.15 |
| Comoros | 0.452 | Iceland | 0.256 | Mongolia | 2.071 | **South Africa** | **36.85** | | |

Table 2. The snapshotted population of nations (1990).

| Country | IHDIGDP | Country | IHDIGDP | Country | IHDIGDP | Country | IHDIGDP | Country | IHDIGDP |
|---|---|---|---|---|---|---|---|---|---|
| Afghanistan | 20.41 | Congo, DRC | 26.21 | **India** | **1,892** | Montenegro | NaN | Spain | 407 |
| Albania | 12.04 | Congo | 6.02 | **Indonesia** | **519.8** | Morocco | 66.14 | Sri Lanka | 52.22 |
| Algeria | 85.45 | Costa Rica | 18.44 | Iran | 250.7 | Mozambique | 12.03 | St. Kitts and Nevis | NaN |
| Angola | 32.25 | Ivory | 18.25 | Iraq | NaN | Myanmar | 50.46 | St. Lucia | NaN |
| Antigua and Barbuda | NaN | Croatia | 27.33 | Ireland | 48.26 | Namibia | 4.656 | St. Vincent and the Grenadines | NaN |
| Argentina | 199.8 | Cyprus | 6.496 | Israel | 58.18 | Nepal | 26.77 | Sudan | 39.33 |
| Armenia | 10.53 | Czech Republic | 84.32 | Italy | 533.6 | The Netherlands | 189 | Suriname | 1.776 |
| **Australia** | **240.7** | Denmark | 57.98 | Jamaica | 10.02 | New Zealand | 31.41 | Swaziland | 2.3 |
| Austria | 92.7 | Djibouti | 0.885 | **Japan** | **1,120** | Nicaragua | 11.79 | Sweden | 98.58 |
| Azerbaijan | 34.94 | Dominica | NaN | Jordan | 13.98 | Niger | 6.873 | **Switzerland** | **88.64** |
| The Bahamas | 2.548 | Dominican Republic | 30.21 | Kazakhstan | 72.22 | Nigeria | 153.7 | Syria | 42.54 |
| Bahrain | 7.012 | Ecuador | 44.98 | Kenya | 40.44 | Norway | 68.92 | Taiwan | NaN |
| Bangladesh | 185.5 | Egypt | 183.8 | Kiribati | NaN | Oman | NaN | Tajikistan | 12.04 |
| Barbados | 1.752 | El Salvador | 17.97 | South Korea | 411.4 | Pakistan | 223.9 | Tanzania | NaN |
| Belarus | 48.7 | Equatorial Guinea | 5.747 | Kosovo | NaN | Panama | 13.58 | Thailand | 220.2 |
| Belgium | 112.7 | Eritrea | NaN | Kuwait | 32.58 | Papua New Guinea | 6.121 | East Timor | 1.535 |
| Belize | 0.898 | Estonia | 8.684 | Kyrgyzstan | 10.54 | Paraguay | 13.17 | Togo | 5.208 |
| Benin | 8.214 | Ethiopia | 53.48 | Laos | 8.616 | Peru | 93.53 | Tonga | 0.288 |
| Bhutan | NaN | Fiji | 1.77 | Latvia | 12.58 | Philippines | 182.6 | Trinidad and Tobago | NaN |
| Bolivia | 19.07 | Finland | 53.72 | Lebanon | NaN | Poland | 246.9 | Tunisia | 35.63 |
| Bosnia and Herzegovina | 13.81 | France | 621.4 | Lesotho | 2.388 | Portugal | 77.91 | Turkey | 289.4 |
| Botswana | 7.304 | Gabon | 6.326 | Liberia | 2.327 | Qatar | 29.14 | Turkmenistan | 14.01 |
| **Brazil** | **702.8** | The Gambia | 1.434 | Libya | 25.36 | Romania | 111.9 | Tuvalu | NaN |
| Brunei | 4.697 | Georgia | 13.32 | Lithuania | 20.35 | **Russia** | **796** | Uganda | 26.25 |
| Bulgaria | 41.25 | **Germany** | **848.8** | Luxembourg | 9.779 | Rwanda | 8.483 | Ukraine | 167.8 |
| Burkina Faso | 9.923 | Ghana | 30.16 | Macedonia | 8.263 | Samoa | NaN | United Arab Emirates | 55.48 |
| Burundi | 4.02 | Greece | 100.5 | Madagascar | 17.31 | Sao Tome and Principe | 0.172 | United Kingdom | 624.9 |
| Cambodia | 16.55 | Grenada | NaN | Malawi | 11.6 | Saudi Arabia | 153.6 | **United States** | **3,832** |
| Cameroon | 21.58 | Guatemala | 25.62 | Malaysia | 110.6 | Senegal | 12.28 | Uruguay | 16.62 |
| **Canada** | **362.2** | Guinea | 6.786 | Maldives | 0.928 | Serbia | 34.17 | Uzbekistan | 55.91 |
| Cape Verde | 0.747 | Guinea-Bissau | 0.941 | Mali | 8.785 | Seychelles | NaN | Vanuatu | NaN |
| Central African Republic | 2.59 | Guyana | 2.431 | Malta | 2.758 | Sierra Leone | 3.334 | Venezuela | NaN |
| Chad | 7.139 | Haiti | 8.644 | Mauritania | 3.302 | Singapore | 61.16 | Vietnam | 165.9 |
| Chile | 82.74 | Honduras | 14.33 | Mauritius | 5.226 | Slovakia | 39.52 | Yemen | 24.88 |
| **China** | **4,026** | Hong Kong | 76.51 | Mexico | 497.9 | Slovenia | 17.61 | Zambia | 11.4 |
| Colombia | 150.1 | Hungary | 67.37 | Moldova | NaN | Solomon Islands | 0.64 | Zimbabwe | 4.025 |
| Comoros | 0.549 | Iceland | 3.419 | Mongolia | 6.054 | **South Africa** | **164.8** | | |

Table 3. The IHDIGDP of nations (in B$) in 2009.

| Country | Emissions (MtCO2e) | Country | Emissions (MtCO2e) | Country | Emissions (MtCO2e) | Country | Emissions (MtCO2e) | Country | Emissions (MtCO2e) |
|---|---|---|---|---|---|---|---|---|---|
| Afghanistan | NaN | Congo, DRC | 113.7 | **India** | **2,432** | Montenegro | NaN | Spain | 405.3 |
| Albania | 8.107 | Congo | 15.86 | **Indonesia** | **812.9** | Morocco | 53.77 | Sri Lanka | 26.24 |
| Algeria | 169.7 | Costa Rica | 10.84 | Iran | 669.8 | Mozambique | 24.87 | St. Kitts and Nevis | NaN |
| Angola | 111.7 | Ivory | 24.99 | Iraq | 117.8 | Myanmar | NaN | St. Lucia | NaN |
| Antigua and Barbuda | NaN | Croatia | 28.41 | Ireland | 64.78 | Namibia | 13.28 | St. Vincent and the Grenadines | NaN |
| Argentina | 329.8 | Cyprus | 10.69 | Israel | 78.54 | Nepal | 31.31 | Sudan | 139.1 |
| Armenia | 15.84 | Czech Republic | 117.1 | Italy | 494.7 | The Netherlands | 288.5 | Suriname | NaN |
| **Australia** | **601.6** | Denmark | 65.67 | Jamaica | 14.15 | New Zealand | 83.35 | Swaziland | NaN |
| Austria | 85.33 | Djibouti | NaN | **Japan** | **1,238** | Nicaragua | 14.21 | Sweden | 70.55 |
| Azerbaijan | 71.45 | Dominica | NaN | Jordan | 23.13 | Niger | NaN | **Switzerland** | **55.72** |
| The Bahamas | NaN | Dominican Republic | 28.91 | Kazakhstan | 265.9 | Nigeria | 254.4 | Syria | 77.99 |
| Bahrain | 34.5 | Ecuador | 54.18 | Kenya | 48.48 | Norway | 66.16 | Taiwan | NaN |
| Bangladesh | 173.2 | Egypt | NaN | Kiribati | NaN | Oman | 72.89 | Tajikistan | 12.77 |
| Barbados | NaN | El Salvador | 10.88 | South Korea | 582.3 | Pakistan | 324.9 | Tanzania | 67.82 |
| Belarus | 85.79 | Equatorial Guinea | NaN | Kosovo | NaN | Panama | 20.54 | Thailand | 368.4 |
| Belgium | 157.1 | Eritrea | 3.87 | Kuwait | 105.8 | Papua New Guinea | NaN | East Timor | NaN |
| Belize | NaN | Estonia | 20.77 | Kyrgyzstan | 10.85 | Paraguay | 30.7 | Togo | 7.645 |
| Benin | 10.45 | Ethiopia | 98.39 | Laos | NaN | Peru | 63.3 | Tonga | NaN |
| Bhutan | NaN | Fiji | NaN | Latvia | 14.81 | Philippines | 140.1 | Trinidad and Tobago | 54.83 |
| Bolivia | 74.38 | Finland | 70.13 | Lebanon | 16.67 | Poland | 394.8 | Tunisia | 34.12 |
| Bosnia and Herzegovina | 22.01 | France | 541.9 | Lesotho | NaN | Portugal | 75.47 | Turkey | 363 |
| Botswana | 14.12 | Gabon | 14.31 | Liberia | NaN | Qatar | 86.75 | Turkmenistan | 96.81 |
| **Brazil** | **1,385** | The Gambia | NaN | Libya | 74.64 | Romania | 118.5 | Tuvalu | NaN |
| Brunei | 15.74 | Georgia | 11.91 | Lithuania | 24.81 | **Russia** | **2,361** | Uganda | NaN |
| Bulgaria | 60.18 | **Germany** | **940.5** | Luxembourg | 12.27 | Rwanda | NaN | Ukraine | 353.8 |
| Burkina Faso | NaN | Ghana | 22.2 | Macedonia | 9.521 | Samoa | NaN | United Arab Emirates | 216.5 |
| Burundi | NaN | Greece | 115.9 | Madagascar | NaN | Sao Tome and Principe | NaN | United Kingdom | 629.6 |
| Cambodia | 38.3 | Grenada | NaN | Malawi | NaN | Saudi Arabia | 534.6 | **United States** | **6,581** |
| Cameroon | 33.25 | Guatemala | 19.26 | Malaysia | 219.6 | Senegal | 17.89 | Uruguay | 36.3 |
| **Canada** | **711.7** | Guinea | NaN | Maldives | NaN | Serbia | 71.57 | Uzbekistan | 169.4 |
| Cape Verde | NaN | Guinea-Bissau | NaN | Mali | NaN | Seychelles | NaN | Vanuatu | NaN |
| Central African Republic | NaN | Guyana | NaN | Malta | 3.547 | Sierra Leone | NaN | Venezuela | NaN |
| Chad | NaN | Haiti | 7.548 | Mauritania | NaN | Singapore | 167.8 | Vietnam | 216.6 |
| Chile | 93.44 | Honduras | 17.81 | Mauritius | NaN | Slovakia | 43.37 | Yemen | 34.32 |
| **China** | **10,060** | Hong Kong | 89.35 | Mexico | 625.8 | Slovenia | 22.56 | Zambia | 55.85 |
| Colombia | 152 | Hungary | 66.39 | Moldova | 11.98 | Solomon Islands | NaN | Zimbabwe | 25.75 |
| Comoros | NaN | Iceland | 4.41 | Mongolia | 16.92 | **South Africa** | **550.6** | | |

Table 4. The GHG emissions of nations in 2009.

| **Country** | **IHDI** | **Country** | **IHDI** | **Country** | **IHDI** | **Country** | **IHDI** | **Country** | **IHDI** |
|---|---|---|---|---|---|---|---|---|---|
| Afghanistan | 1,867 | Congo, DRC | 1,492 | **India** | **3,601** | Montenegro | 6,921 | Spain | 7,772 |
| Albania | 6,244 | Congo | 3,299 | **Indonesia** | **4,882** | Morocco | 4,035 | Sri Lanka | 5,419 |
| Algeria | 3,662 | Costa Rica | 5,744 | Iran | 3,804 | Mozambique | 1,529 | St. Kitts and Nevis | NaN |
| Angola | 2,395 | Ivory | 2,521 | Iraq | NaN | Myanmar | 2,423 | St. Lucia | NaN |
| Antigua and Barbuda | NaN | Croatia | 6,483 | Ireland | 8,121 | Namibia | 3,363 | St. Vincent and the Grenadines | NaN |
| Argentina | 6,196 | Cyprus | 7,151 | Israel | 7,621 | Nepal | 2,886 | Sudan | 2,047 |
| Armenia | 6,172 | Czech Republic | 7,900 | Italy | 7,494 | The Netherlands | 8,162 | Suriname | 4,867 |
| **Australia** | **8,622** | Denmark | 8,082 | Jamaica | 5,724 | New Zealand | 4,934 | Swaziland | 3,161 |
| Austria | 7,852 | Djibouti | 2,501 | **Japan** | **4,808** | Nicaragua | 4,238 | Sweden | 8,231 |
| Azerbaijan | 6,114 | Dominica | NaN | Jordan | 5,468 | Niger | 1,710 | **Switzerland** | **8,111** |
| The Bahamas | 6,701 | Dominican Republic | 4,967 | Kazakhstan | 6,144 | Nigeria | 2,437 | Syria | 4,646 |
| Bahrain | 4,355 | Ecuador | 5,517 | Kenya | 3,159 | Norway | 8,751 | Taiwan | NaN |
| Bangladesh | 3,268 | Egypt | 4,447 | Kiribati | NaN | Oman | NaN | Tajikistan | 4,658 |
| Barbados | 4,295 | El Salvador | 4,741 | South Korea | 7,268 | Pakistan | 3,339 | Tanzania | NaN |
| Belarus | 6,613 | Equatorial Guinea | 2,925 | Kosovo | NaN | Panama | 5,381 | Thailand | 5,113 |
| Belgium | 7,922 | Eritrea | NaN | Kuwait | 4,197 | Papua New Guinea | 2,325 | East Timor | 3,307 |
| Belize | 4,950 | Estonia | 7,303 | Kyrgyzstan | 5,046 | Paraguay | 4,775 | Togo | 2,850 |
| Benin | 2,801 | Ethiopia | 2,134 | Laos | 3,687 | Peru | 4,975 | Tonga | 3,684 |
| Bhutan | NaN | Fiji | 3,640 | Latvia | 6,841 | Philippines | 5,156 | Trinidad and Tobago | 6,176 |
| Bolivia | 3,943 | Finland | 8,042 | Lebanon | NaN | Poland | 7,054 | Tunisia | 5,065 |
| Bosnia and Herzegovina | 5,642 | France | 7,893 | Lesotho | 2,794 | Portugal | 6,965 | Turkey | 5,141 |
| Botswana | 3,422 | Gabon | 5,073 | Liberia | 1,842 | Qatar | 4,355 | Turkmenistan | 4,878 |
| **Brazil** | **5,046** | The Gambia | 2,349 | Libya | 4,088 | Romania | 6,724 | Tuvalu | NaN |
| Brunei | 4,389 | Georgia | 5,765 | Lithuania | 6,921 | **Russia** | **6,316** | Uganda | 2,819 |
| Bulgaria | 6,572 | **Germany** | **8,122** | Luxembourg | 7,732 | Rwanda | 2,392 | Ukraine | 6,483 |
| Burkina Faso | 1,937 | Ghana | 3,460 | Macedonia | 5,807 | Samoa | NaN | United Arab Emirates | 4,432 |
| Burundi | 1,733 | Greece | 7,662 | Madagascar | 3,087 | Sao Tome and Principe | 2,647 | United Kingdom | 7,642 |
| Cambodia | 3,474 | Grenada | NaN | Malawi | 2,549 | Saudi Arabia | 4,082 | **United States** | **7,963** |
| Cameroon | 3,014 | Guatemala | 3,693 | Malaysia | 4,033 | Senegal | 2,601 | Uruguay | 6,378 |
| **Canada** | **8,103** | Guinea | 2,077 | Maldives | 5,021 | Serbia | 6,542 | Uzbekistan | 5,168 |
| Cape Verde | 2,898 | Guinea-Bissau | 1,643 | Mali | 1,885 | Seychelles | NaN | Vanuatu | NaN |
| Central African Republic | 1,807 | Guyana | 4,921 | Malta | 4,437 | Sierra Leone | 1,906 | Venezuela | NaN |
| Chad | 1,778 | Haiti | 2,425 | Mauritania | 2,785 | Singapore | 4,590 | Vietnam | 4,730 |
| Chile | 6,308 | Honduras | 4,169 | Mauritius | 3,804 | Slovakia | 7,612 | Yemen | 2,837 |
| **China** | **5,048** | Hong Kong | 4,677 | Mexico | 5,890 | Slovenia | 7,691 | Zambia | 2,645 |
| Colombia | 4,892 | Hungary | 7,342 | Moldova | NaN | Solomon Islands | 2,685 | Zimbabwe | 825.4 |
| Comoros | 2,389 | Iceland | 8,111 | Mongolia | 5,219 | **South Africa** | **4,089** | | |

Table 5. The IHDI of the nations in 2009.

| Country | IHDIxCapita | Country | IHDIxCapita | Country | IHDIxCapita | Country | IHDIxCapita | Country | IHDIxCapita |
|---|---|---|---|---|---|---|---|---|---|
| Afghanistan | 42,570 | Congo, DRC | 61,540 | **India** | **3,104,000** | Montenegro | NaN | Spain | 301,800 |
| Albania | 19,960 | Congo | 7,363 | **Indonesia** | **878,000** | Morocco | 97,010 | Sri Lanka | 88,140 |
| Algeria | 91,630 | Costa Rica | 21,890 | Iran | 207,300 | Mozambique | 21,630 | St. Kitts and Nevis | NaN |
| Angola | 25,830 | Ivory | 29,540 | Iraq | NaN | Myanmar | 98,840 | St. Lucia | NaN |
| Antigua and Barbuda | NaN | Croatia | 28,400 | Ireland | 28,470 | Namibia | 4,524 | St. Vincent and the Grenadines | NaN |
| Argentina | 201,600 | Cyprus | 4,141 | Israel | 34,400 | Nepal | 55,160 | Sudan | 52,700 |
| Armenia | 19,820 | Czech Republic | 80,600 | Italy | 424,800 | The Netherlands | 122,000 | Suriname | 2,244 |
| **Australia** | **148,000** | Denmark | 41,500 | Jamaica | 13,580 | New Zealand | 16,830 | Swaziland | 2,839 |
| Austria | 60,280 | Djibouti | 1,268 | **Japan** | **593,500** | Nicaragua | 23,120 | Sweden | 70,520 |
| Azerbaijan | 45,610 | Dominica | NaN | Jordan | 18,960 | Niger | 13,220 | **Switzerland** | **54,440** |
| The Bahamas | 1,702 | Dominican Republic | 35,510 | Kazakhstan | 91,250 | Nigeria | 220,700 | Syria | 56,330 |
| Bahrain | 2,091 | Ecuador | 58,650 | Kenya | 76,270 | Norway | 37,190 | Taiwan | NaN |
| Bangladesh | 377,800 | Egypt | 228,400 | Kiribati | NaN | Oman | NaN | Tajikistan | 25,780 |
| Barbados | 1,074 | El Salvador | 24,230 | South Korea | 311,600 | Pakistan | 362,000 | Tanzania | NaN |
| Belarus | 62,690 | Equatorial Guinea | 1,322 | Kosovo | NaN | Panama | 12,880 | Thailand | 287,900 |
| Belgium | 79,030 | Eritrea | NaN | Kuwait | 8,940 | Papua New Guinea | 8,737 | East Timor | 2,586 |
| Belize | 935.6 | Estonia | 9,786 | Kyrgyzstan | 22,540 | Paraguay | 19,520 | Togo | 11,290 |
| Benin | 14,930 | Ethiopia | 103,000 | Laos | 15,510 | Peru | 108,200 | Tonga | 353.7 |
| Bhutan | NaN | Fiji | 2,636 | Latvia | 15,470 | Philippines | 317,100 | Trinidad and Tobago | NaN |
| Bolivia | 25,680 | Finland | 40,200 | Lebanon | NaN | Poland | 269,500 | Tunisia | 41,300 |
| Bosnia and Herzegovina | 20,620 | France | 447,600 | Lesotho | 4,813 | Portugal | 69,620 | Turkey | 269,600 |
| Botswana | 4,380 | Gabon | 4,718 | Liberia | 5,658 | Qatar | 1,838 | Turkmenistan | 18,840 |
| **Brazil** | **739,800** | The Gambia | 2,028 | Libya | 17,840 | Romania | 158,300 | Tuvalu | NaN |
| Brunei | 1,110 | Georgia | 24,880 | Lithuania | 23,120 | **Russia** | **932,800** | Uganda | 49,990 |
| Bulgaria | 57,300 | **Germany** | **640,700** | Luxembourg | 2,954 | Rwanda | 17,120 | Ukraine | 296,300 |
| Burkina Faso | 16,530 | Ghana | 49,500 | Macedonia | 11,160 | Samoa | NaN | United Arab Emirates | 8,173 |
| Burundi | 9,462 | Greece | 77,850 | Madagascar | 37,150 | Sao Tome and Principe | 304.4 | United Kingdom | 437,400 |
| Cambodia | 29,490 | Grenada | NaN | Malawi | 25,590 | Saudi Arabia | 62,000 | **United States** | **1,991,000** |
| Cameroon | 34,730 | Guatemala | 30,330 | Malaysia | 73,010 | Senegal | 20,750 | Uruguay | 19,730 |
| **Canada** | **223,900** | Guinea | 12,770 | Maldives | 1,064 | Serbia | 48,290 | Uzbekistan | 110,400 |
| Cape Verde | 999.8 | Guinea-Bissau | 1,610 | Mali | 15,060 | Seychelles | NaN | Vanuatu | NaN |
| Central African Republic | 5,321 | Guyana | 3,696 | Malta | 1,571 | Sierra Leone | 6,723 | Venezuela | NaN |
| Chad | 10,040 | Haiti | 17,160 | Mauritania | 5,456 | Singapore | 14,390 | Vietnam | 312,300 |
| Chile | 82,630 | Honduras | 20,430 | Mauritius | 4,028 | Slovakia | 40,400 | Yemen | 34,290 |
| **China** | **5,772,000** | Hong Kong | 26,900 | Mexico | 490,200 | Slovenia | 15,210 | Zambia | 20,920 |
| Colombia | 166,900 | Hungary | 76,170 | Moldova | NaN | Solomon Islands | 862 | Zimbabwe | 8,382 |
| Comoros | 1,080 | Iceland | 2,076 | Mongolia | 10,810 | **South Africa** | **150,700** | | |

Table 6. The IHDIxCapita of the nation in 2009.

| Country | GHG-INT | Country | GHG-INT | Country | GHG-INT | Country | GHG-INT | Country | GHG-INT |
|---|---|---|---|---|---|---|---|---|---|
| Afghanistan | NaN | Congo, DRC | 5.327 | **India** | **0.667** | Montenegro | NaN | Spain | 0.298 |
| Albania | 0.355 | Congo | 1.021 | **Indonesia** | **0.846** | Morocco | 0.37 | Sri Lanka | 0.271 |
| Algeria | 0.705 | Costa Rica | 0.223 | Iran | 0.834 | Mozambique | 1.232 | St. Kitts and Nevis | NaN |
| Angola | 1.068 | Ivory | 0.699 | Iraq | 1.058 | Myanmar | NaN | St. Lucia | NaN |
| Antigua and Barbuda | NaN | Croatia | 0.362 | Ireland | 0.376 | Namibia | 0.959 | St. Vincent and the Grenadines | NaN |
| Argentina | 0.566 | Cyprus | 0.47 | Israel | 0.378 | Nepal | 0.923 | Sudan | 1.475 |
| Armenia | 0.973 | Czech Republic | 0.463 | Italy | 0.285 | The Netherlands | 0.438 | Suriname | NaN |
| **Australia** | **0.707** | Denmark | 0.335 | Jamaica | 0.596 | New Zealand | 0.725 | Swaziland | NaN |
| Austria | 0.265 | Djibouti | NaN | **Japan** | **0.302** | Nicaragua | 0.846 | Sweden | 0.212 |
| Azerbaijan | 0.834 | Dominica | NaN | Jordan | 0.697 | Niger | NaN | **Switzerland** | **0.178** |
| The Bahamas | NaN | Dominican Republic | 0.36 | Kazakhstan | 1.463 | Nigeria | 0.737 | Syria | 0.757 |
| Bahrain | 1.22 | Ecuador | 0.491 | Kenya | 0.778 | Norway | 0.263 | Taiwan | NaN |
| Bangladesh | 0.717 | Egypt | NaN | Kiribati | NaN | Oman | 1.011 | Tajikistan | 0.932 |
| Barbados | NaN | El Salvador | 0.254 | South Korea | 0.428 | Pakistan | 0.74 | Tanzania | 1.248 |
| Belarus | 0.71 | Equatorial Guinea | NaN | Kosovo | NaN | Panama | 0.503 | Thailand | 0.683 |
| Belgium | 0.41 | Eritrea | 1.101 | Kuwait | 0.798 | Papua New Guinea | NaN | East Timor | NaN |
| Belize | NaN | Estonia | 0.876 | Kyrgyzstan | 0.899 | Paraguay | 1.073 | Togo | 1.336 |
| Benin | 0.773 | Ethiopia | 1.246 | Laos | NaN | Peru | 0.252 | Tonga | NaN |
| Bhutan | NaN | Fiji | NaN | Latvia | 0.458 | Philippines | 0.432 | Trinidad and Tobago | 2.122 |
| Bolivia | 1.634 | Finland | 0.393 | Lebanon | 0.305 | Poland | 0.574 | Tunisia | 0.357 |
| Bosnia and Herzegovina | 0.738 | France | 0.259 | Lesotho | NaN | Portugal | 0.313 | Turkey | 0.413 |
| Botswana | 0.543 | Gabon | 0.679 | Liberia | NaN | Qatar | 0.676 | Turkmenistan | 2.892 |
| **Brazil** | **0.692** | The Gambia | NaN | Libya | 0.867 | Romania | 0.465 | Tuvalu | NaN |
| Brunei | 0.812 | Georgia | 0.57 | Lithuania | 0.448 | **Russia** | **1.115** | Uganda | NaN |
| Bulgaria | 0.629 | **Germany** | **0.334** | Luxembourg | 0.312 | Rwanda | NaN | Ukraine | 1.22 |
| Burkina Faso | NaN | Ghana | 0.382 | Macedonia | 0.484 | Samoa | NaN | United Arab Emirates | 0.914 |
| Burundi | NaN | Greece | 0.351 | Madagascar | NaN | Sao Tome and Principe | NaN | United Kingdom | 0.296 |
| Cambodia | 1.358 | Grenada | NaN | Malawi | NaN | Saudi Arabia | 0.9 | **United States** | **0.466** |
| Cameroon | 0.78 | Guatemala | 0.284 | Malaysia | 0.573 | Senegal | 0.788 | Uruguay | 0.828 |
| **Canada** | **0.557** | Guinea | NaN | Maldives | NaN | Serbia | 0.918 | Uzbekistan | 2.162 |
| Cape Verde | NaN | Guinea-Bissau | NaN | Mali | NaN | Seychelles | NaN | Vanuatu | NaN |
| Central African Republic | NaN | Guyana | NaN | Malta | 0.357 | Sierra Leone | NaN | Venezuela | NaN |
| Chad | NaN | Haiti | 0.63 | Mauritania | NaN | Singapore | 0.664 | Vietnam | 0.844 |
| Chile | 0.385 | Honduras | 0.55 | Mauritius | NaN | Slovakia | 0.379 | Yemen | 0.59 |
| **China** | **1.111** | Hong Kong | 0.296 | Mexico | 0.425 | Slovenia | 0.407 | Zambia | 3.027 |
| Colombia | 0.368 | Hungary | 0.362 | Moldova | 1.177 | Solomon Islands | NaN | Zimbabwe | 5.192 |
| Comoros | NaN | Iceland | 0.364 | Mongolia | 1.645 | **South Africa** | **1.09** | | |

Table 7. The GHG-INT of the nation in 2009.

| Country | MGHG-INT | Country | MGHG-INT | Country | MGHG-INT | Country | MGHG-INT | Country | MGHG-INT |
|---|---|---|---|---|---|---|---|---|---|
| Afghanistan | NaN | Congo, DRC | 4.34 | **India** | **1.29** | Montenegro | NaN | Spain | 1 |
| Albania | 0.67 | Congo | 2.64 | **Indonesia** | **1.56** | Morocco | 0.81 | Sri Lanka | 0.5 |
| Algeria | 1.99 | Costa Rica | 0.59 | Iran | 2.67 | Mozambique | 2.07 | St. Kitts and Nevis | NaN |
| Angola | 3.46 | Ivory | 1.37 | Iraq | NaN | Myanmar | NaN | St. Lucia | NaN |
| Antigua and Barbuda | NaN | Croatia | 1.04 | Ireland | 1.34 | Namibia | 2.85 | St. Vincent and the Grenadines | NaN |
| Argentina | 1.65 | Cyprus | 1.65 | Israel | 1.35 | Nepal | 1.17 | Sudan | 3.54 |
| Armenia | 1.5 | Czech Republic | 1.39 | Italy | 0.93 | The Netherlands | 1.53 | Suriname | NaN |
| **Australia** | **2.5** | Denmark | 1.13 | Jamaica | 1.41 | New Zealand | 2.65 | Swaziland | NaN |
| Austria | 0.92 | Djibouti | NaN | **Japan** | **1.11** | Nicaragua | 1.2 | Sweden | 0.72 |
| Azerbaijan | 2.05 | Dominica | NaN | Jordan | 1.65 | Niger | NaN | **Switzerland** | **0.63** |
| The Bahamas | NaN | Dominican Republic | 0.96 | Kazakhstan | 3.68 | Nigeria | 1.66 | Syria | 1.83 |
| Bahrain | 4.92 | Ecuador | 1.2 | Kenya | 1.2 | Norway | 0.96 | Taiwan | NaN |
| Bangladesh | 0.93 | Egypt | NaN | Kiribati | NaN | Oman | NaN | Tajikistan | 1.06 |
| Barbados | NaN | El Salvador | 0.61 | South Korea | 1.42 | Pakistan | 1.45 | Tanzania | NaN |
| Belarus | 1.76 | Equatorial Guinea | NaN | Kosovo | NaN | Panama | 1.51 | Thailand | 1.67 |
| Belgium | 1.39 | Eritrea | NaN | Kuwait | 3.25 | Papua New Guinea | NaN | East Timor | NaN |
| Belize | NaN | Estonia | 2.39 | Kyrgyzstan | 1.03 | Paraguay | 2.33 | Togo | 1.47 |
| Benin | 1.27 | Ethiopia | 1.84 | Laos | NaN | Peru | 0.68 | Tonga | NaN |
| Bhutan | NaN | Fiji | NaN | Latvia | 1.18 | Philippines | 0.77 | Trinidad and Tobago | NaN |
| Bolivia | 3.9 | Finland | 1.31 | Lebanon | NaN | Poland | 1.6 | Tunisia | 0.96 |
| Bosnia and Herzegovina | 1.59 | France | 0.87 | Lesotho | NaN | Portugal | 0.97 | Turkey | 1.25 |
| Botswana | 1.93 | Gabon | 2.26 | Liberia | NaN | Qatar | 2.98 | Turkmenistan | 6.91 |
| **Brazil** | **1.97** | The Gambia | NaN | Libya | 2.94 | Romania | 1.06 | Tuvalu | NaN |
| Brunei | 3.35 | Georgia | 0.89 | Lithuania | 1.22 | **Russia** | **2.97** | Uganda | NaN |
| Bulgaria | 1.46 | **Germany** | **1.11** | Luxembourg | 1.25 | Rwanda | NaN | Ukraine | 2.11 |
| Burkina Faso | NaN | Ghana | 0.74 | Macedonia | 1.15 | Samoa | NaN | United Arab Emirates | 3.9 |
| Burundi | NaN | Greece | 1.15 | Madagascar | NaN | Sao Tome and Principe | NaN | United Kingdom | 1.01 |
| Cambodia | 2.31 | Grenada | NaN | Malawi | NaN | Saudi Arabia | 3.48 | **United States** | **1.72** |
| Cameroon | 1.54 | Guatemala | 0.75 | Malaysia | 1.99 | Senegal | 1.46 | Uruguay | 2.18 |
| **Canada** | **1.96** | Guinea | NaN | Maldives | NaN | Serbia | 2.09 | Uzbekistan | 3.03 |
| Cape Verde | NaN | Guinea-Bissau | NaN | Mali | NaN | Seychelles | NaN | Vanuatu | NaN |
| Central African Republic | NaN | Guyana | NaN | Malta | 1.29 | Sierra Leone | NaN | Venezuela | NaN |
| Chad | NaN | Haiti | 0.87 | Mauritania | NaN | Singapore | 2.74 | Vietnam | 1.31 |
| Chile | 1.13 | Honduras | 1.24 | Mauritius | NaN | Slovakia | 1.1 | Yemen | 1.38 |
| **China** | **2.5** | Hong Kong | 1.17 | Mexico | 1.26 | Slovenia | 1.28 | Zambia | 4.9 |
| Colombia | 1.01 | Hungary | 0.99 | Moldova | NaN | Solomon Islands | NaN | Zimbabwe | 6.4 |
| Comoros | NaN | Iceland | 1.29 | Mongolia | 2.79 | **South Africa** | **3.34** | | |

Table 8. The MGHG-INT of the nation in 2009.

| **Country** | **Admissible Emissions (MtCO2e)** | **Country** | **Admissible Emissions (MtCO2e)** | **Country** | **Admissible Emissions (MtCO2e)** | **Country** | **Admissible Emissions (MtCO2e)** | **Country** | **Admissible Emissions (MtCO2e)** |
|---|---|---|---|---|---|---|---|---|---|
| Afghanistan | 30.2 | Congo, DRC | 38.8 | **India** | **2,801** | Montenegro | NaN | Spain | 602.3 |
| Albania | 17.8 | Congo | 8.9 | **Indonesia** | **769.3** | Morocco | 97.9 | Sri Lanka | 77.3 |
| Algeria | 126.5 | Costa Rica | 27.3 | Iran | 371 | Mozambique | 17.8 | St. Kitts and Nevis | NaN |
| Angola | 47.7 | Ivory | 27 | Iraq | NaN | Myanmar | 74.7 | St. Lucia | NaN |
| Antigua and Barbuda | NaN | Croatia | 40.5 | Ireland | 71.4 | Namibia | 6.9 | St. Vincent and the Grenadines | NaN |
| Argentina | 295.7 | Cyprus | 9.6 | Israel | 86.1 | Nepal | 39.6 | Sudan | 58.2 |
| Armenia | 15.6 | Czech Republic | 124.8 | Italy | 789.6 | The Netherlands | 279.7 | Suriname | 2.6 |
| **Australia** | **356.2** | Denmark | 85.8 | Jamaica | 14.8 | New Zealand | 46.5 | Swaziland | 3.4 |
| Austria | 137.2 | Djibouti | 1.3 | **Japan** | **1,657** | Nicaragua | 17.5 | Sweden | 145.9 |
| Azerbaijan | 51.7 | Dominica | NaN | Jordan | 20.7 | Niger | 10.2 | **Switzerland** | **131.2** |
| The Bahamas | 3.8 | Dominican Republic | 44.7 | Kazakhstan | 106.9 | Nigeria | 227.5 | Syria | 63 |
| Bahrain | 10.4 | Ecuador | 66.6 | Kenya | 59.9 | Norway | 102 | Taiwan | NaN |
| Bangladesh | 274.5 | Egypt | 272.1 | Kiribati | NaN | Oman | NaN | Tajikistan | 17.8 |
| Barbados | 2.6 | El Salvador | 26.6 | South Korea | 608.8 | Pakistan | 331.4 | Tanzania | NaN |
| Belarus | 72.1 | Equatorial Guinea | 8.5 | Kosovo | NaN | Panama | 20.1 | Thailand | 325.9 |
| Belgium | 166.8 | Eritrea | NaN | Kuwait | 48.2 | Papua New Guinea | 9.1 | East Timor | 2.3 |
| Belize | 1.3 | Estonia | 12.9 | Kyrgyzstan | 15.6 | Paraguay | 19.5 | Togo | 7.7 |
| Benin | 12.2 | Ethiopia | 79.2 | Laos | 12.8 | Peru | 138.4 | Tonga | 0.4 |
| Bhutan | NaN | Fiji | 2.6 | Latvia | 18.6 | Philippines | 270.3 | Trinidad and Tobago | NaN |
| Bolivia | 28.2 | Finland | 79.5 | Lebanon | NaN | Poland | 365.4 | Tunisia | 52.7 |
| Bosnia and Herzegovina | 20.4 | France | 919.7 | Lesotho | 3.5 | Portugal | 115.3 | Turkey | 428.3 |
| Botswana | 10.8 | Gabon | 9.4 | Liberia | 3.4 | Qatar | 43.1 | Turkmenistan | 20.7 |
| **Brazil** | **1,040** | The Gambia | 2.1 | Libya | 37.5 | Romania | 165.5 | Tuvalu | NaN |
| Brunei | 7 | Georgia | 19.7 | Lithuania | 30.1 | **Russia** | **1,178** | Uganda | 38.9 |
| Bulgaria | 61.1 | **Germany** | **1,256** | Luxembourg | 14.5 | Rwanda | 12.6 | Ukraine | 248.3 |
| Burkina Faso | 14.7 | Ghana | 44.6 | Macedonia | 12.2 | Samoa | NaN | United Arab Emirates | 82.1 |
| Burundi | 5.9 | Greece | 148.7 | Madagascar | 25.6 | Sao Tome and Principe | 0.3 | United Kingdom | 924.8 |
| Cambodia | 24.5 | Grenada | NaN | Malawi | 17.2 | Saudi Arabia | 227.3 | **United States** | **5,671** |
| Cameroon | 31.9 | Guatemala | 37.9 | Malaysia | 163.7 | Senegal | 18.2 | Uruguay | 24.6 |
| **Canada** | **536** | Guinea | 10 | Maldives | 1.4 | Serbia | 50.6 | Uzbekistan | 82.7 |
| Cape Verde | 1.1 | Guinea-Bissau | 1.4 | Mali | 13 | Seychelles | NaN | Vanuatu | NaN |
| Central African Republic | 3.8 | Guyana | 3.6 | Malta | 4.1 | Sierra Leone | 4.9 | Venezuela | NaN |
| Chad | 10.6 | Haiti | 12.8 | Mauritania | 4.9 | Singapore | 90.5 | Vietnam | 245.5 |
| Chile | 122.5 | Honduras | 21.2 | Mauritius | 7.7 | Slovakia | 58.5 | Yemen | 36.8 |
| **China** | **5,958** | Hong Kong | 113.2 | Mexico | 737 | Slovenia | 26.1 | Zambia | 16.9 |
| Colombia | 222.1 | Hungary | 99.7 | Moldova | NaN | Solomon Islands | 0.9 | Zimbabwe | 6 |
| Comoros | 0.8 | Iceland | 5.1 | Mongolia | 9 | **South Africa** | **243.8** | | |

Table 9. The admissible emissions of nations in 2009.

| **Country** | **Admissible Emissions per Capita (tCO2e)** | **Country** | **Admissible Emissions per Capita (tCO2e)** | **Country** | **Admissible Emissions per Capita (tCO2e)** | **Country** | **Admissible Emissions per Capita (tCO2e)** | **Country** | **Admissible Emissions per Capita (tCO2e)** |
|---|---|---|---|---|---|---|---|---|---|
| Afghanistan | 1.32 | Congo, DRC | 0.94 | **India** | **3.25** | Montenegro | NaN | Spain | 15.51 |
| Albania | 5.57 | Congo | 3.99 | **Indonesia** | **4.28** | Morocco | 4.07 | Sri Lanka | 4.75 |
| Algeria | 5.06 | Costa Rica | 7.17 | Iran | 6.81 | Mozambique | 1.26 | St. Kitts and Nevis | NaN |
| Angola | 4.42 | Ivory | 2.3 | Iraq | NaN | Myanmar | 1.83 | St. Lucia | NaN |
| Antigua and Barbuda | NaN | Croatia | 9.24 | Ireland | 20.37 | Namibia | 5.13 | St. Vincent and the Grenadines | NaN |
| Argentina | 9.09 | Cyprus | 16.58 | Israel | 19.07 | Nepal | 2.07 | Sudan | 2.26 |
| Armenia | 4.86 | Czech Republic | 12.23 | Italy | 13.93 | The Netherlands | 18.71 | Suriname | 5.64 |
| **Australia** | **20.75** | Denmark | 16.71 | Jamaica | 6.24 | New Zealand | 13.63 | Swaziland | 3.79 |
| Austria | 17.87 | Djibouti | 2.56 | **Japan** | **13.42** | Nicaragua | 3.21 | Sweden | 17.03 |
| Azerbaijan | 6.93 | Dominica | NaN | Jordan | 5.97 | Niger | 1.32 | **Switzerland** | **19.55** |
| The Bahamas | 14.96 | Dominican Republic | 6.25 | Kazakhstan | 7.2 | Nigeria | 2.51 | Syria | 5.2 |
| Bahrain | 21.67 | Ecuador | 6.27 | Kenya | 2.48 | Norway | 24 | Taiwan | NaN |
| Bangladesh | 2.37 | Egypt | 5.3 | Kiribati | NaN | Oman | NaN | Tajikistan | 3.22 |
| Barbados | 10.4 | El Salvador | 5.21 | South Korea | 14.2 | Pakistan | 3.06 | Tanzania | NaN |
| Belarus | 7.61 | Equatorial Guinea | 18.81 | Kosovo | NaN | Panama | 8.4 | Thailand | 5.79 |
| Belgium | 16.72 | Eritrea | NaN | Kuwait | 22.63 | Papua New Guinea | 2.42 | East Timor | 2.94 |
| Belize | 6.88 | Estonia | 9.63 | Kyrgyzstan | 3.49 | Paraguay | 4.77 | Togo | 1.94 |
| Benin | 2.29 | Ethiopia | 1.64 | Laos | 3.04 | Peru | 6.36 | Tonga | 4.17 |
| Bhutan | NaN | Fiji | 3.59 | Latvia | 8.23 | Philippines | 4.4 | Trinidad and Tobago | NaN |
| Bolivia | 4.33 | Finland | 15.91 | Lebanon | NaN | Poland | 9.57 | Tunisia | 6.46 |
| Bosnia and Herzegovina | 5.58 | France | 16.22 | Lesotho | 2.03 | Portugal | 11.53 | Turkey | 8.17 |
| Botswana | 8.44 | Gabon | 10.11 | Liberia | 1.11 | Qatar | 102.1 | Turkmenistan | 5.36 |
| **Brazil** | **7.1** | The Gambia | 2.43 | Libya | 8.59 | Romania | 7.03 | Tuvalu | NaN |
| Brunei | 27.67 | Georgia | 4.57 | Lithuania | 9.01 | **Russia** | **7.98** | Uganda | 2.19 |
| Bulgaria | 7.01 | **Germany** | **15.92** | Luxembourg | 37.96 | Rwanda | 1.76 | Ukraine | 5.43 |
| Burkina Faso | 1.72 | Ghana | 3.12 | Macedonia | 6.35 | Samoa | NaN | United Arab Emirates | 44.52 |
| Burundi | 1.08 | Greece | 14.63 | Madagascar | 2.13 | Sao Tome and Principe | 2.61 | United Kingdom | 16.16 |
| Cambodia | 2.89 | Grenada | NaN | Malawi | 1.71 | Saudi Arabia | 14.97 | **United States** | **22.68** |
| Cameroon | 2.77 | Guatemala | 4.62 | Malaysia | 9.04 | Senegal | 2.28 | Uruguay | 7.95 |
| **Canada** | **19.4** | Guinea | 1.63 | Maldives | 6.6 | Serbia | 6.85 | Uzbekistan | 3.87 |
| Cape Verde | 3.19 | Guinea-Bissau | 1.43 | Mali | 1.63 | Seychelles | NaN | Vanuatu | NaN |
| Central African Republic | 1.29 | Guyana | 4.79 | Malta | 11.58 | Sierra Leone | 1.39 | Venezuela | NaN |
| Chad | 1.88 | Haiti | 1.81 | Mauritania | 2.5 | Singapore | 28.87 | Vietnam | 3.72 |
| Chile | 9.35 | Honduras | 4.33 | Mauritius | 7.27 | Slovakia | 11.02 | Yemen | 3.04 |
| **China** | **5.21** | Hong Kong | 19.68 | Mexico | 8.86 | Slovenia | 13.2 | Zambia | 2.14 |
| Colombia | 6.51 | Hungary | 9.61 | Moldova | NaN | Solomon Islands | 2.8 | Zimbabwe | 0.59 |
| Comoros | 1.77 | Iceland | 19.92 | Mongolia | 4.35 | **South Africa** | **6.62** | | |

Table 10. The admissible emissions per capita of nations in 2009.

| Country | Emissions Debt (MtCO2e) | Country | Emissions Debt (MtCO2e) | Country | Emissions Debt (MtCO2e) | Country | Emissions Debt (MtCO2e) | Country | Emissions Debt (MtCO2e) |
|---|---|---|---|---|---|---|---|---|---|
| Afghanistan | NaN | Congo, DRC | 74.94 | **India** | **0** | Montenegro | NaN | Spain | 0 |
| Albania | 0 | Congo | 6.965 | **Indonesia** | **43.58** | Morocco | 0 | Sri Lanka | 0 |
| Algeria | 43.17 | Costa Rica | 0 | Iran | 298.8 | Mozambique | 7.073 | St. Kitts and Nevis | NaN |
| Angola | 64.01 | Ivory | 0 | Iraq | NaN | Myanmar | NaN | St. Lucia | NaN |
| Antigua and Barbuda | NaN | Croatia | 0 | Ireland | 0 | Namibia | 6.378 | St. Vincent and the Grenadines | NaN |
| Argentina | 34.14 | Cyprus | 1.092 | Israel | 0 | Nepal | 0 | Sudan | 80.89 |
| Armenia | 0.242 | Czech Republic | 0 | Italy | 0 | The Netherlands | 8.814 | Suriname | NaN |
| **Australia** | **245.4** | Denmark | 0 | Jamaica | 0 | New Zealand | 36.85 | Swaziland | NaN |
| Austria | 0 | Djibouti | NaN | **Japan** | **0** | Nicaragua | 0 | Sweden | 0 |
| Azerbaijan | 19.75 | Dominica | NaN | Jordan | 2.433 | Niger | NaN | **Switzerland** | **0** |
| The Bahamas | NaN | Dominican Republic | 0 | Kazakhstan | 159 | Nigeria | 26.9 | Syria | 14.99 |
| Bahrain | 24.1 | Ecuador | 0 | Kenya | 0 | Norway | 0 | Taiwan | NaN |
| Bangladesh | 0 | Egypt | NaN | Kiribati | NaN | Oman | NaN | Tajikistan | 0 |
| Barbados | NaN | El Salvador | 0 | South Korea | 0 | Pakistan | 0 | Tanzania | NaN |
| Belarus | 13.69 | Equatorial Guinea | NaN | Kosovo | NaN | Panama | 0.44 | Thailand | 42.54 |
| Belgium | 0 | Eritrea | NaN | Kuwait | 57.61 | Papua New Guinea | NaN | East Timor | NaN |
| Belize | NaN | Estonia | 7.873 | Kyrgyzstan | 0 | Paraguay | 11.2 | Togo | 0 |
| Benin | 0 | Ethiopia | 19.19 | Laos | NaN | Peru | 0 | Tonga | NaN |
| Bhutan | NaN | Fiji | NaN | Latvia | 0 | Philippines | 0 | Trinidad and Tobago | NaN |
| Bolivia | 46.18 | Finland | 0 | Lebanon | NaN | Poland | 29.42 | Tunisia | 0 |
| Bosnia and Herzegovina | 1.612 | France | 0 | Lesotho | NaN | Portugal | 0 | Turkey | 0 |
| Botswana | 3.318 | Gabon | 4.91 | Liberia | NaN | Qatar | 43.65 | Turkmenistan | 76.11 |
| **Brazil** | **345.1** | The Gambia | NaN | Libya | 37.14 | Romania | 0 | Tuvalu | NaN |
| Brunei | 8.74 | Georgia | 0 | Lithuania | 0 | **Russia** | **1,183** | Uganda | NaN |
| Bulgaria | 0 | **Germany** | **0** | Luxembourg | 0 | Rwanda | NaN | Ukraine | 105.5 |
| Burkina Faso | NaN | Ghana | 0 | Macedonia | 0 | Samoa | NaN | United Arab Emirates | 134.4 |
| Burundi | NaN | Greece | 0 | Madagascar | NaN | Sao Tome and Principe | NaN | United Kingdom | 0 |
| Cambodia | 13.8 | Grenada | NaN | Malawi | NaN | Saudi Arabia | 307.3 | **United States** | **909.4** |
| Cameroon | 1.353 | Guatemala | 0 | Malaysia | 55.87 | Senegal | 0 | Uruguay | 11.7 |
| **Canada** | **175.7** | Guinea | NaN | Maldives | NaN | Serbia | 20.97 | Uzbekistan | 86.75 |
| Cape Verde | NaN | Guinea-Bissau | NaN | Mali | NaN | Seychelles | NaN | Vanuatu | NaN |
| Central African Republic | NaN | Guyana | NaN | Malta | 0 | Sierra Leone | NaN | Venezuela | NaN |
| Chad | NaN | Haiti | 0 | Mauritania | NaN | Singapore | 77.32 | Vietnam | 0 |
| Chile | 0 | Honduras | 0 | Mauritius | NaN | Slovakia | 0 | Yemen | 0 |
| **China** | **4,105** | Hong Kong | 0 | Mexico | 0 | Slovenia | 0 | Zambia | 38.95 |
| Colombia | 0 | Hungary | 0 | Moldova | NaN | Solomon Islands | NaN | Zimbabwe | 19.75 |
| Comoros | NaN | Iceland | 0 | Mongolia | 7.915 | **South Africa** | **306.8** | | |

Table 11. The Emission debt of nations in 2009.

| Country | Emissions Credit (MtCO2e) | Country | Emissions Credit (MtCO2e) | Country | Emissions Credit (MtCO2e) | Country | Emissions Credit (MtCO2e) | Country | Emissions Credit (MtCO2e) |
|---|---|---|---|---|---|---|---|---|---|
| Afghanistan | NaN | Congo, DRC | 0 | **India** | **368.7** | Montenegro | NaN | Spain | 197 |
| Albania | 9.693 | Congo | 0 | **Indonesia** | **0** | Morocco | 44.13 | Sri Lanka | 51.06 |
| Algeria | 0 | Costa Rica | 16.46 | Iran | 0 | Mozambique | 0 | St. Kitts and Nevis | NaN |
| Angola | 0 | Ivory | 2.01 | Iraq | NaN | Myanmar | NaN | St. Lucia | NaN |
| Antigua and Barbuda | NaN | Croatia | 12.09 | Ireland | 6.623 | Namibia | 0 | St. Vincent and the Grenadines | NaN |
| Argentina | 0 | Cyprus | 0 | Israel | 7.559 | Nepal | 8.295 | Sudan | 0 |
| Armenia | 0 | Czech Republic | 7.73 | Italy | 294.9 | The Netherlands | 0 | Suriname | NaN |
| **Australia** | **0** | Denmark | 20.13 | Jamaica | 0.65 | New Zealand | 0 | Swaziland | NaN |
| Austria | 51.87 | Djibouti | NaN | **Japan** | **418.8** | Nicaragua | 3.292 | Sweden | 75.35 |
| Azerbaijan | 0 | Dominica | NaN | Jordan | 0 | Niger | NaN | **Switzerland** | **75.48** |
| The Bahamas | NaN | Dominican Republic | 15.79 | Kazakhstan | 0 | Nigeria | 0 | Syria | 0 |
| Bahrain | 0 | Ecuador | 12.42 | Kenya | 11.42 | Norway | 35.84 | Taiwan | NaN |
| Bangladesh | 101.3 | Egypt | NaN | Kiribati | NaN | Oman | NaN | Tajikistan | 5.026 |
| Barbados | NaN | El Salvador | 15.72 | South Korea | 26.46 | Pakistan | 6.457 | Tanzania | NaN |
| Belarus | 0 | Equatorial Guinea | NaN | Kosovo | NaN | Panama | 0 | Thailand | 0 |
| Belgium | 9.673 | Eritrea | NaN | Kuwait | 0 | Papua New Guinea | NaN | East Timor | NaN |
| Belize | NaN | Estonia | 0 | Kyrgyzstan | 4.749 | Paraguay | 0 | Togo | 0.055 |
| Benin | 1.748 | Ethiopia | 0 | Laos | NaN | Peru | 75.1 | Tonga | NaN |
| Bhutan | NaN | Fiji | NaN | Latvia | 3.788 | Philippines | 130.2 | Trinidad and Tobago | NaN |
| Bolivia | 0 | Finland | 9.37 | Lebanon | NaN | Poland | 0 | Tunisia | 18.58 |
| Bosnia and Herzegovina | 0 | France | 377.8 | Lesotho | NaN | Portugal | 39.83 | Turkey | 65.33 |
| Botswana | 0 | Gabon | 0 | Liberia | NaN | Qatar | 0 | Turkmenistan | 0 |
| **Brazil** | **0** | The Gambia | NaN | Libya | 0 | Romania | 46.98 | Tuvalu | NaN |
| Brunei | 0 | Georgia | 7.788 | Lithuania | 5.295 | **Russia** | **0** | Uganda | NaN |
| Bulgaria | 0.923 | **Germany** | **315.7** | Luxembourg | 2.235 | Rwanda | NaN | Ukraine | 0 |
| Burkina Faso | NaN | Ghana | 22.4 | Macedonia | 2.679 | Samoa | NaN | United Arab Emirates | 0 |
| Burundi | NaN | Greece | 32.85 | Madagascar | NaN | Sao Tome and Principe | NaN | United Kingdom | 295.2 |
| Cambodia | 0 | Grenada | NaN | Malawi | NaN | Saudi Arabia | 0 | **United States** | **0** |
| Cameroon | 0 | Guatemala | 18.64 | Malaysia | 0 | Senegal | 0.308 | Uruguay | 0 |
| **Canada** | **0** | Guinea | NaN | Maldives | NaN | Serbia | 0 | Uzbekistan | 0 |
| Cape Verde | NaN | Guinea-Bissau | NaN | Mali | NaN | Seychelles | NaN | Vanuatu | NaN |
| Central African Republic | NaN | Guyana | NaN | Malta | 0.553 | Sierra Leone | NaN | Venezuela | NaN |
| Chad | NaN | Haiti | 5.252 | Mauritania | NaN | Singapore | 0 | Vietnam | 28.91 |
| Chile | 29.06 | Honduras | 3.389 | Mauritius | NaN | Slovakia | 15.13 | Yemen | 2.483 |
| **China** | **0** | Hong Kong | 23.85 | Mexico | 111.2 | Slovenia | 3.538 | Zambia | 0 |
| Colombia | 70.1 | Hungary | 33.31 | Moldova | NaN | Solomon Islands | NaN | Zimbabwe | 0 |
| Comoros | NaN | Iceland | 0.69 | Mongolia | 0 | **South Africa** | **0** | | |

Table 12. The Emission credit of nations in 2009.

| Country | RED % | Country | RED % | Country | RED % | Country | RED % | Country | RED % |
|---|---|---|---|---|---|---|---|---|---|
| Afghanistan | NaN | Congo, DRC | 1,896 | **India** | **0** | Montenegro | NaN | Spain | 0 |
| Albania | 0 | Congo | 768 | **Indonesia** | **56** | Morocco | 0 | Sri Lanka | 0 |
| Algeria | 335 | Costa Rica | 0 | Iran | 790 | Mozambique | 390 | St. Kitts and Nevis | NaN |
| Angola | 1,317 | Ivory | 0 | Iraq | NaN | Myanmar | NaN | St. Lucia | NaN |
| Antigua and Barbuda | NaN | Croatia | 0 | Ireland | 0 | Namibia | 907 | St. Vincent and the Grenadines | NaN |
| Argentina | 113 | Cyprus | 112 | Israel | 0 | Nepal | 0 | Sudan | 1,364 |
| Armenia | 15 | Czech Republic | 0 | Italy | 0 | The Netherlands | 31 | Suriname | NaN |
| **Australia** | **676** | Denmark | 0 | Jamaica | 0 | New Zealand | 778 | Swaziland | NaN |
| Austria | 0 | Djibouti | NaN | **Japan** | **0** | Nicaragua | 0 | Sweden | 0 |
| Azerbaijan | 375 | Dominica | NaN | Jordan | 115 | Niger | NaN | **Switzerland** | **0** |
| The Bahamas | NaN | Dominican Republic | 0 | Kazakhstan | 1,460 | Nigeria | 116 | Syria | 234 |
| Bahrain | 2,274 | Ecuador | 0 | Kenya | 0 | Norway | 0 | Taiwan | NaN |
| Bangladesh | 0 | Egypt | NaN | Kiribati | NaN | Oman | NaN | Tajikistan | 0 |
| Barbados | NaN | El Salvador | 0 | South Korea | 0 | Pakistan | 0 | Tanzania | NaN |
| Belarus | 186 | Equatorial Guinea | NaN | Kosovo | NaN | Panama | 21 | Thailand | 128 |
| Belgium | 0 | Eritrea | NaN | Kuwait | 1,173 | Papua New Guinea | NaN | East Timor | NaN |
| Belize | NaN | Estonia | 599 | Kyrgyzstan | 0 | Paraguay | 564 | Togo | 0 |
| Benin | 0 | Ethiopia | 238 | Laos | NaN | Peru | 0 | Tonga | NaN |
| Bhutan | NaN | Fiji | NaN | Latvia | 0 | Philippines | 0 | Trinidad and Tobago | NaN |
| Bolivia | 1,607 | Finland | 0 | Lebanon | NaN | Poland | 79 | Tunisia | 0 |
| Bosnia and Herzegovina | 78 | France | 0 | Lesotho | NaN | Portugal | 0 | Turkey | 0 |
| Botswana | 302 | Gabon | 513 | Liberia | NaN | Qatar | 994 | Turkmenistan | 3,608 |
| **Brazil** | **326** | The Gambia | NaN | Libya | 972 | Romania | 0 | Tuvalu | NaN |
| Brunei | 1,225 | Georgia | 0 | Lithuania | 0 | **Russia** | **986** | Uganda | NaN |
| Bulgaria | 0 | **Germany** | **0** | Luxembourg | 0 | Rwanda | NaN | Ukraine | 417 |
| Burkina Faso | NaN | Ghana | 0 | Macedonia | 0 | Samoa | NaN | United Arab Emirates | 1,607 |
| Burundi | NaN | Greece | 0 | Madagascar | NaN | Sao Tome and Principe | NaN | United Kingdom | 0 |
| Cambodia | 553 | Grenada | NaN | Malawi | NaN | Saudi Arabia | 1,327 | **United States** | **157** |
| Cameroon | 42 | Guatemala | 0 | Malaysia | 335 | Senegal | 0 | Uruguay | 467 |
| **Canada** | **322** | Guinea | NaN | Maldives | NaN | Serbia | 407 | Uzbekistan | 1,029 |
| Cape Verde | NaN | Guinea-Bissau | NaN | Mali | NaN | Seychelles | NaN | Vanuatu | NaN |
| Central African Republic | NaN | Guyana | NaN | Malta | 0 | Sierra Leone | NaN | Venezuela | NaN |
| Chad | NaN | Haiti | 0 | Mauritania | NaN | Singapore | 839 | Vietnam | 0 |
| Chile | 0 | Honduras | 0 | Mauritius | NaN | Slovakia | 0 | Yemen | 0 |
| **China** | **676** | Hong Kong | 0 | Mexico | 0 | Slovenia | 0 | Zambia | 2,262 |
| Colombia | 0 | Hungary | 0 | Moldova | NaN | Solomon Islands | NaN | Zimbabwe | 3,230 |
| Comoros | NaN | Iceland | 0 | Mongolia | 863 | **South Africa** | **1,235** | | |

Table 13. The RED percentage of nations in 2009. The $1/\text{RED}_{\text{BCT}} = 1/100$ of these values are used as the border carbon tax of 2010.

| Country | Emissions (MtCO2e) | Country | Emissions (MtCO2e) | Country | Emissions (MtCO2e) | Country | Emissions (MtCO2e) | Country | Emissions (MtCO2e) |
|---|---|---|---|---|---|---|---|---|---|
| Afghanistan | NaN | Congo, DRC | 2,759 | **India** | **35,200** | Montenegro | NaN | Spain | 7,381 |
| Albania | 146.5 | Congo | 256.5 | **Indonesia** | **11,420** | Morocco | 912 | Sri Lanka | 447.4 |
| Algeria | 2,932 | Costa Rica | 193.3 | Iran | 9,515 | Mozambique | 458.1 | St. Kitts and Nevis | NaN |
| Angola | 1,954 | Ivory | 499.4 | Iraq | 1,954 | Myanmar | NaN | St. Lucia | NaN |
| Antigua and Barbuda | NaN | Croatia | 546.6 | Ireland | 1,209 | Namibia | 190.2 | St. Vincent and the Grenadines | NaN |
| Argentina | 5,686 | Cyprus | 170.7 | Israel | 1,265 | Nepal | 557.1 | Sudan | 2,201 |
| Armenia | 252.3 | Czech Republic | 2,412 | Italy | 10,440 | The Netherlands | 5,758 | Suriname | NaN |
| **Australia** | **10,720** | Denmark | 1,514 | Jamaica | 249.6 | New Zealand | 1,503 | Swaziland | NaN |
| Austria | 1,645 | Djibouti | NaN | **Japan** | **25,860** | Nicaragua | 241 | Sweden | 1,584 |
| Azerbaijan | 1,422 | Dominica | NaN | Jordan | 349.8 | Niger | NaN | **Switzerland** | **1,081** |
| The Bahamas | NaN | Dominican Republic | 457.5 | Kazakhstan | 4,570 | Nigeria | 4,886 | Syria | 1,299 |
| Bahrain | 477.8 | Ecuador | 829.9 | Kenya | 771.5 | Norway | 1,317 | Taiwan | NaN |
| Bangladesh | 2,782 | Egypt | NaN | Kiribati | NaN | Oman | 723.6 | Tajikistan | 246.5 |
| Barbados | NaN | El Salvador | 184.4 | South Korea | 9,431 | Pakistan | 4,909 | Tanzania | 1,068 |
| Belarus | 1,749 | Equatorial Guinea | NaN | Kosovo | NaN | Panama | 351.6 | Thailand | 5,613 |
| Belgium | 3,222 | Eritrea | 86.54 | Kuwait | 1,372 | Papua New Guinea | NaN | East Timor | NaN |
| Belize | NaN | Estonia | 428.7 | Kyrgyzstan | 243.8 | Paraguay | 552.3 | Togo | 120 |
| Benin | 187.9 | Ethiopia | 1,556 | Laos | NaN | Peru | 1,004 | Tonga | NaN |
| Bhutan | NaN | Fiji | NaN | Latvia | 274.9 | Philippines | 2,483 | Trinidad and Tobago | 795.1 |
| Bolivia | 987.9 | Finland | 1,430 | Lebanon | 295.6 | Poland | 8,400 | Tunisia | 543.9 |
| Bosnia and Herzegovina | 383.6 | France | 10,880 | Lesotho | NaN | Portugal | 1,464 | Turkey | 5,684 |
| Botswana | 244.8 | Gabon | 274.1 | Liberia | NaN | Qatar | 948.6 | Turkmenistan | 1,243 |
| **Brazil** | **18,870** | The Gambia | NaN | Libya | 1,250 | Romania | 3,031 | Tuvalu | NaN |
| Brunei | 225.7 | Georgia | 238.5 | Lithuania | 496.9 | **Russia** | **46,310** | Uganda | NaN |
| Bulgaria | 1,396 | **Germany** | **20,900** | Luxembourg | 245.5 | Rwanda | NaN | Ukraine | 9,461 |
| Burkina Faso | NaN | Ghana | 376.7 | Macedonia | 222.1 | Samoa | NaN | United Arab Emirates | 2,931 |
| Burundi | NaN | Greece | 2,220 | Madagascar | NaN | Sao Tome and Principe | NaN | United Kingdom | 14,190 |
| Cambodia | 492.6 | Grenada | NaN | Malawi | NaN | Saudi Arabia | 7,186 | **United States** | **133,300** |
| Cameroon | 724.7 | Guatemala | 489.2 | Malaysia | 3,398 | Senegal | 287.3 | Uruguay | 624.2 |
| **Canada** | **14,060** | Guinea | NaN | Maldives | NaN | Serbia | 1,427 | Uzbekistan | 3,177 |
| Cape Verde | NaN | Guinea-Bissau | NaN | Mali | NaN | Seychelles | NaN | Vanuatu | NaN |
| Central African Republic | NaN | Guyana | NaN | Malta | 62.2 | Sierra Leone | NaN | Venezuela | NaN |
| Chad | NaN | Haiti | 126 | Mauritania | NaN | Singapore | 2,266 | Vietnam | 2,869 |
| Chile | 1,503 | Honduras | 247.4 | Mauritius | NaN | Slovakia | 958.1 | Yemen | 447.3 |
| **China** | **111,600** | Hong Kong | 1,275 | Mexico | 10,620 | Slovenia | 420.7 | Zambia | 1,011 |
| Colombia | 2,675 | Hungary | 1,496 | Moldova | 262.3 | Solomon Islands | NaN | Zimbabwe | 567 |
| Comoros | NaN | Iceland | 79.78 | Mongolia | 395.8 | **South Africa** | **9,364** | | |

Table 14. The cumulative GHG emissions of nations from 1990 to 2009.

| Country | CIHDIGDP | Country | CIHDIGDP | Country | CIHDIGDP | Country | CIHDIGDP | Country | CIHDIGDP |
|---|---|---|---|---|---|---|---|---|---|
| Afghanistan | 323.9 | Congo, DRC | 427.5 | **India** | **27,280** | Montenegro | NaN | Spain | 8,186 |
| Albania | 191.8 | Congo | 106 | **Indonesia** | **8,111** | Morocco | 1,068 | Sri Lanka | 846.3 |
| Algeria | 1,528 | Costa Rica | 316.1 | Iran | 4,449 | Mozambique | 163.9 | St. Kitts and Nevis | NaN |
| Angola | 442.3 | Ivory | 337.9 | Iraq | NaN | Myanmar | 709.9 | St. Lucia | NaN |
| Antigua and Barbuda | NaN | Croatia | 550.6 | Ireland | 934 | Namibia | 81.67 | St. Vincent and the Grenadines | NaN |
| Argentina | 3,554 | Cyprus | 124.8 | Israel | 1,088 | Nepal | 402.3 | Sudan | 585.9 |
| Armenia | 169.3 | Czech Republic | 1,805 | Italy | 12,220 | The Netherlands | 4,002 | Suriname | 30.9 |
| **Australia** | **4,668** | Denmark | 1,293 | Jamaica | 203 | New Zealand | 656 | Swaziland | 43.65 |
| Austria | 1,977 | Djibouti | 16.14 | **Japan** | **26,840** | Nicaragua | 191.3 | Sweden | 2,094 |
| Azerbaijan | 461.1 | Dominica | NaN | Jordan | 220.7 | Niger | 103.2 | **Switzerland** | **1,930** |
| The Bahamas | 55.04 | Dominican Republic | 498.4 | Kazakhstan | 1,198 | Nigeria | 2,397 | Syria | 733.7 |
| Bahrain | 118 | Ecuador | 785.9 | Kenya | 673.9 | Norway | 1,458 | Taiwan | NaN |
| Bangladesh | 2,621 | Egypt | 2,967 | Kiribati | NaN | Oman | NaN | Tajikistan | 186.5 |
| Barbados | 37.46 | El Salvador | 327.7 | South Korea | 7,345 | Pakistan | 3,566 | Tanzania | NaN |
| Belarus | 806.4 | Equatorial Guinea | 60.62 | Kosovo | NaN | Panama | 216.2 | Thailand | 3,890 |
| Belgium | 2,414 | Eritrea | NaN | Kuwait | 625 | Papua New Guinea | 109.9 | East Timor | 26.54 |
| Belize | 15.99 | Estonia | 178.7 | Kyrgyzstan | 170.9 | Paraguay | 238.1 | Togo | 83.92 |
| Benin | 127 | Ethiopia | 751 | Laos | 122.8 | Peru | 1,544 | Tonga | 5.754 |
| Bhutan | NaN | Fiji | 34.03 | Latvia | 245.9 | Philippines | 2,976 | Trinidad and Tobago | NaN |
| Bolivia | 333.1 | Finland | 1,123 | Lebanon | NaN | Poland | 4,278 | Tunisia | 586.4 |
| Bosnia and Herzegovina | 272.5 | France | 13,390 | Lesotho | 40.09 | Portugal | 1,656 | Turkey | 5,290 |
| Botswana | 132.1 | Gabon | 135.5 | Liberia | 36.65 | Qatar | 322.8 | Turkmenistan | 206.9 |
| **Brazil** | **13,210** | The Gambia | 23.92 | Libya | 512.2 | Romania | 1,978 | Tuvalu | NaN |
| Brunei | 110.7 | Georgia | 228 | Lithuania | 420.4 | **Russia** | **15,000** | Uganda | 360.3 |
| Bulgaria | 743.8 | **Germany** | **18,990** | Luxembourg | 192.6 | Rwanda | 109.3 | Ukraine | 3,158 |
| Burkina Faso | 160.6 | Ghana | 461.4 | Macedonia | 154.1 | Samoa | NaN | United Arab Emirates | 972.5 |
| Burundi | 60.19 | Greece | 1,967 | Madagascar | 283 | Sao Tome and Principe | 2.777 | United Kingdom | 13,330 |
| Cambodia | 247.2 | Grenada | NaN | Malawi | 170.7 | Saudi Arabia | 3,041 | **United States** | **81,850** |
| Cameroon | 373.2 | Guatemala | 459.6 | Malaysia | 1,968 | Senegal | 199.6 | Uruguay | 298.2 |
| **Canada** | **7,579** | Guinea | 117.5 | Maldives | 14.11 | Serbia | 700.6 | Uzbekistan | 884.3 |
| Cape Verde | 11.91 | Guinea-Bissau | 17.55 | Mali | 127.6 | Seychelles | NaN | Vanuatu | NaN |
| Central African Republic | 44.6 | Guyana | 41.85 | Malta | 68.92 | Sierra Leone | 48.74 | Venezuela | NaN |
| Chad | 118.8 | Haiti | 156.3 | Mauritania | 54.67 | Singapore | 1,084 | Vietnam | 2,358 |
| Chile | 1,492 | Honduras | 247.3 | Mauritius | 93.28 | Slovakia | 750.6 | Yemen | 430.5 |
| **China** | **52,890** | Hong Kong | 1,492 | Mexico | 9,973 | Slovenia | 360.5 | Zambia | 187.5 |
| Colombia | 2,683 | Hungary | 1,326 | Moldova | NaN | Solomon Islands | 11.97 | Zimbabwe | 120 |
| Comoros | 9.75 | Iceland | 68.55 | Mongolia | 90.89 | **South Africa** | **3,107** | | |

Table 15. The cumulative IHDIGDP calculated from 1990 to 2009 (in B$).

| Country | REDC % | Country | REDC % | Country | REDC % | Country | REDC % | Country | REDC % |
|---|---|---|---|---|---|---|---|---|---|
| Afghanistan | NaN | Congo, DRC | 6,712 | **India** | **0** | Montenegro | NaN | Spain | 0 |
| Albania | 0 | Congo | 1,219 | **Indonesia** | **0** | Morocco | 0 | Sri Lanka | 0 |
| Algeria | 537 | Costa Rica | 0 | Iran | 837 | Mozambique | 1,731 | St. Kitts and Nevis | NaN |
| Angola | 3,939 | Ivory | 0 | Iraq | NaN | Myanmar | NaN | St. Lucia | NaN |
| Antigua and Barbuda | NaN | Croatia | 0 | Ireland | 0 | Namibia | 1,096 | St. Vincent and the Grenadines | NaN |
| Argentina | 104 | Cyprus | 0 | Israel | 0 | Nepal | 0 | Sudan | 3,039 |
| Armenia | 0 | Czech Republic | 0 | Italy | 0 | The Netherlands | 0 | Suriname | NaN |
| **Australia** | **1,052** | Denmark | 0 | Jamaica | 0 | New Zealand | 1,045 | Swaziland | NaN |
| Austria | 0 | Djibouti | NaN | **Japan** | **0** | Nicaragua | 0 | Sweden | 0 |
| Azerbaijan | 2,125 | Dominica | NaN | Jordan | 83 | Niger | NaN | **Switzerland** | **0** |
| The Bahamas | NaN | Dominican Republic | 0 | Kazakhstan | 3,119 | Nigeria | 700 | Syria | 335 |
| Bahrain | 3,435 | Ecuador | 0 | Kenya | 0 | Norway | 0 | Taiwan | NaN |
| Bangladesh | 0 | Egypt | NaN | Kiribati | NaN | Oman | NaN | Tajikistan | 0 |
| Barbados | NaN | El Salvador | 0 | South Korea | 0 | Pakistan | 0 | Tanzania | NaN |
| Belarus | 879 | Equatorial Guinea | NaN | Kosovo | NaN | Panama | 139 | Thailand | 0 |
| Belgium | 0 | Eritrea | NaN | Kuwait | 913 | Papua New Guinea | NaN | East Timor | NaN |
| Belize | NaN | Estonia | 1,192 | Kyrgyzstan | 0 | Paraguay | 1,083 | Togo | 0 |
| Benin | 0 | Ethiopia | 747 | Laos | NaN | Peru | 0 | Tonga | NaN |
| Bhutan | NaN | Fiji | NaN | Latvia | 0 | Philippines | 0 | Trinidad and Tobago | NaN |
| Bolivia | 1,962 | Finland | 0 | Lebanon | NaN | Poland | 598 | Tunisia | 0 |
| Bosnia and Herzegovina | 0 | France | 0 | Lesotho | NaN | Portugal | 0 | Turkey | 0 |
| Botswana | 448 | Gabon | 677 | Liberia | NaN | Qatar | 1,926 | Turkmenistan | 6,101 |
| **Brazil** | **0** | The Gambia | NaN | Libya | 1,246 | Romania | 11 | Tuvalu | NaN |
| Brunei | 700 | Georgia | 0 | Lithuania | 0 | **Russia** | **2,128** | Uganda | NaN |
| Bulgaria | 480 | **Germany** | **0** | Luxembourg | 0 | Rwanda | NaN | Ukraine | 2,004 |
| Burkina Faso | NaN | Ghana | 0 | Macedonia | 0 | Samoa | NaN | United Arab Emirates | 2,029 |
| Burundi | NaN | Greece | 0 | Madagascar | NaN | Sao Tome and Principe | NaN | United Kingdom | 0 |
| Cambodia | 639 | Grenada | NaN | Malawi | NaN | Saudi Arabia | 1,142 | **United States** | **143** |
| Cameroon | 569 | Guatemala | 0 | Malaysia | 276 | Senegal | 0 | Uruguay | 775 |
| **Canada** | **450** | Guinea | NaN | Maldives | NaN | Serbia | 699 | Uzbekistan | 2,815 |
| Cape Verde | NaN | Guinea-Bissau | NaN | Mali | NaN | Seychelles | NaN | Vanuatu | NaN |
| Central African Republic | NaN | Guyana | NaN | Malta | 0 | Sierra Leone | NaN | Venezuela | NaN |
| Chad | NaN | Haiti | 0 | Mauritania | NaN | Singapore | 771 | Vietnam | 0 |
| Chile | 0 | Honduras | 0 | Mauritius | NaN | Slovakia | 0 | Yemen | 0 |
| **China** | **798** | Hong Kong | 0 | Mexico | 0 | Slovenia | 0 | Zambia | 5,266 |
| Colombia | 0 | Hungary | 0 | Moldova | NaN | Solomon Islands | NaN | Zimbabwe | 4,357 |
| Comoros | NaN | Iceland | 0 | Mongolia | 3,854 | **South Africa** | **2,029** | | |

Table 16. The cumulative RED percentage calculated based on the cumulative GHG emissions.